\documentclass[12pt]{article}

\usepackage{epsfig}
\input epsf

\title{Prompt photon photoproduction at HERA\\ in the $k_T$-factorization approach}

\author{A.V.~Lipatov, N.P.~Zotov}

\begin{document}

\maketitle

\begin{center}

{\it D.V.~Skobeltsyn Institute of Nuclear Physics,\\ 
M.V. Lomonosov Moscow State University,
\\119992 Moscow, Russia\/}\\[3mm]

\end{center}

\vspace{1cm}

\begin{center}

{\bf Abstract }

\end{center}

We present calculations of the prompt photon photoproduction at HERA collider
in the $k_T$-factorization approach. 
Both direct and resolved
contributions are taken into account. The conservative error analisys
is performed. The unintegrated parton densities in a proton and 
in a photon are determined using the Kimber-Martin-Ryskin prescription. 
We investigate both inclusive and associated with jet prompt photon photoproduction rates.
In particular, we study the angular correlations between produced photon and hadronic jet in the 
transverse momentum plane which can provide a unique information about non-collinear 
evolution dynamics. We compare our theoretical predictions with recent 
experimental data taken by the H1 and ZEUS collaborations.

\vspace{1cm}

\section{Introduction} \indent 

The prompt photon production in $ep$ collisions at HERA is a subject of the 
intensive studies~[1--4]. The theoretical and experimental investigations of the such processes
have provided a direct probe of the hard subprocess dynamics, since produced photons 
are largely insensitive to the effects of final-state hadronization.
Usually photons are called "prompt" if they are coupled to the interacting quarks.
From the theoretical point, these photons in $ep$ collisions can be produced
via direct $\gamma q \to \gamma q$ and resolved production mechanisms. In resolved
events, the photon emitted by the electron fluctuate into a hadronic state
and a gluon and/or a quark of this hadronic fluctuation takes part in the hard
interactions. Also observed final state photon may arise from fragmentation
process~[5], where a quark or gluon decays into $\gamma$. 
The cross section of such processes involves relative poorly known
parton-to-photon fragmentation functions~[6].
However, the isolation 
criterion which introduced in experimental analyses substantially reduces~[7] the 
fragmentation component. 
In any case, for the theoretical description of prompt photon production
at modern (HERA, Tevatron) and 
future (LHC) colliders the detailed knowledge of parton (quark and gluon) distributions
in a proton and in a photon is necessary.

Usually quark and gluon densities are described by the 
Dokshitzer-Gribov-Lipatov-Altarelli-Parizi (DGLAP) evolution equation~[8] 
where large logarithmic terms proportional to $\ln \mu^2$ are taken into 
account only. The cross sections can be rewritten in terms of process-dependent
hard matrix elements convoluted with quark or gluon density functions.
In this way the dominant contributions come from diagrams where parton emissions 
in initial state are strongly ordered in virtuality. This is called collinear 
factorization, as the strong ordering means that the virtuality of the parton 
entering the hard scattering matrix elements can be neglected compared to the 
large scale $\mu$.

However, at high energies (or small $x \sim \mu^2/s \ll 1$) effects of 
finite virtualities and transverse 
momenta of the incoming partons may become more and
more important. These effects can be systematically accounted for in a 
$k_T$-factorization formalism~[9--12]. Just as for DGLAP, in this way it is possible to 
factorize an observable into a convolution of process-dependent 
hard matrix elements with universal parton distributions. 
But as the virtualities and transverse momenta of the emitted partons are no 
longer ordered, the matrix elements have to be taken off-shell and the 
convolution made also over transverse momentum ${\mathbf k}_T$ with the 
unintegrated (i.e.~$k_T$-dependent) parton distributions.
The unintegrated parton distribution $f_a(x,{\mathbf k}_T^2)$ 
determines the probability to find a type $a$ parton carrying the 
longitudinal momentum fraction $x$ and the transverse momentum ${\mathbf k}_T$.
In particular, usage of the unintegrated parton densities have the advantage that it takes
into account true kinematics of the process under consideration even at leading order.

The unintegrated parton distributions $f_a(x,{\mathbf k}_T^2)$ are a subject 
of intensive studies~[13, 14]. Various approaches to investigate these quantities 
has been proposed. It is believed that at assymptotically large energies (or very small $x$) 
the theoretically correct description is given by the Balitsky-Fadin-Kuraev-Lipatov
(BFKL) evolution equation~[15] where large terms proportional to $\ln 1/x$ are 
taken into account. Another approach, valid for both small and large
$x$, have been developed by Ciafaloni, Catani, Fiorani and Marchesini,
and is known as the CCFM model~[16]. It introduces angular ordering of emissions
to correctly treat gluon coherence effects. In the limit of 
asymptotic energies, it almost equivalent to BFKL~[17--19], but also similar to
the DGLAP evolution for large $x \sim 1$. The resulting unintegrated
gluon distribution depends on two scales, the additional scale ${\bar q}$
is a variable related to the maximum angle allowed in the emission and plays 
the role of the evolution scale $\mu$ in the collinear parton densities.

Also it is possible to obtain the two-scale involved unintegrated parton distributions 
from the conventional ones using the Kimber-Martin-Ryskin (KMR) prescription~[20]. 
In this way the $\mu$ dependence in the unintegrated parton distribution 
enters only in last step of the evolution, and single scale evolution equations 
can be used up to this step. Such procedure can be applied to a proton as well as  
photon and is expected to account for the main part of the collinear higher-order QCD corrections. 
The KMR-constructed parton densities were used, in particular, to describe the 
heavy quark production in $\gamma \gamma$ collisions at CERN LEP2~[21] and
prompt photon hadroproduction at fixed target experiments and at 
Fermilab Tevatron~[22] collider (in the double logarithmic approximation).

In the present paper we will apply the KMR method to obtain the unintegrated 
quark and gluon distributions in a proton $f_a(x,{\mathbf k}_T^2,\mu^2)$ and in a photon 
$f_a^\gamma(x,{\mathbf k}_T^2,\mu^2)$ independently from other authors. 
After that, we calculate the inclusive prompt photon 
photoproduction at HERA energies. Such calculations in the $k_T$-factorization 
approach will be performed for the first time. We will investigate the transverse 
energy $E_T^{\gamma}$ and pseudo-rapidity $\eta^{\gamma}$ distributions of the produced 
prompt photons and compare our theoretical results with the recent 
experimental data taken by the H1~[4] and ZEUS~[3] collaborations. 
In order to estimate the theoretical uncertainties of our predictions we will study the 
renormalization and factorization scale dependences of the calculated cross sections.
Next we calculate the associated prompt photon and jet production
rates using some physically motivated approximation. In order
to investigate the underlying dynamics more detail, we will study the angular correlations
between the prompt photon and jet in the transverse momentum plane.
It was shown~[23, 24] that theoretical and experimental studying of such quantities
is a direct probe of the non-collinear parton evolution.

The additional motivation of our investigations within the $k_T$-factorization 
approach is the fact that the next-to-leading order (NLO) 
collinear QCD calculations~[7, 25] are $30 - 40$\% below the data, especially in 
rear pseudo-rapidity (electron direction) region. So, one of the main goals
of this paper is to investigate whether the $k_T$-factorization
formalism could give a better description of the HERA data than collinear NLO QCD calculations.

The our paper is organized as follows. In Section 2 the KMR unintegrated parton densities
in a proton and in a photon are obtained and their properties 
are discussed. In particular, we compare the KMR gluon distributions with ones taken
from the full CCFM equation.
In Section 3 we present the basic
formulas with a brief review of calculation steps.
In Section 4 we present the numerical results of
our calculations. Finally, in Section 5, we give some conclusions.
The compact analytic expressions for the off-mass shell matrix elements of all the 
subprocesses under consideration are given in Appendix. These formulas
may be useful for the subsequent applications.

\section{The KMR unintegrated partons} \indent 

The Kimber-Martin-Ryskin approach~[20] is the formalism to construct
parton distributions $f_a(x,{\mathbf k}_T^2,\mu^2)$ unintegrated over the parton 
transverse momenta ${\mathbf k}_T^2$ from the known conventional parton
distributions $a(x,\mu^2)$, where $a = xg$ or $a = xq$. This formalism is 
valid for a proton as well as photon and can embody
both DGLAP and BFKL contributions. It also accounts for the angular ordering which
comes from coherence effects in gluon emission.

We start from parton distributions in a proton.
The key observation here is that
the $\mu$ dependence of the unintegrated distributions $f_a(x,{\mathbf k}_T^2,\mu^2)$ 
enters at the last step
of the evolution, and therefore single scale evolution equations (DGLAP 
or unified DGLAP-BFKL~[26]) can be used up to this step.
It was shown~[20] that the unintegrated distributions obtained via unified DGLAP-BFKL 
evolution are rather similar to those based on the pure DGLAP equations.
It is because the imposion of the angular ordering constraint is more important than 
including the BFKL effects. Based on this point, 
in our calculations we will use much more simpler DGLAP equation up to the last evolution step.
In this approximation, the unintegrated quark and gluon distributions are given~[20] by
$$
  \displaystyle f_q(x,{\mathbf k}_T^2,\mu^2) = T_q({\mathbf k}_T^2,\mu^2) {\alpha_s({\mathbf k}_T^2)\over 2\pi} \times \atop {
  \displaystyle \times \int\limits_x^1 dz \left[P_{qq}(z) {x\over z} q\left({x\over z},{\mathbf k}_T^2\right) \Theta\left(\Delta - z\right) + P_{qg}(z) {x\over z} g\left({x\over z},{\mathbf k}_T^2\right) \right],} \eqno (1)
$$
$$
  \displaystyle f_g(x,{\mathbf k}_T^2,\mu^2) = T_g({\mathbf k}_T^2,\mu^2) {\alpha_s({\mathbf k}_T^2)\over 2\pi} \times \atop {
  \displaystyle \times \int\limits_x^1 dz \left[\sum_q P_{gq}(z) {x\over z} q\left({x\over z},{\mathbf k}_T^2\right) + P_{gg}(z) {x\over z} g\left({x\over z},{\mathbf k}_T^2\right)\Theta\left(\Delta - z\right) \right],} \eqno (2)
$$

\noindent
where $P_{ab}(z)$ are the usual unregulated leading order DGLAP splitting functions,
and $q(x,\mu^2)$ and $g(x,\mu^2)$ are the conventional quark and gluon densities.
The theta functions which appear in (1) and (2) imply the angular-ordering 
constraint $\Delta = \mu/(\mu + |{\mathbf k}_T|)$ specifically to the last 
evolution step to regulate the soft gluon
singularities. For other evolution steps, the strong ordering in 
transverse momentum within the DGLAP equations automatically 
ensures angular ordering. It is important that parton 
distributions $f_a(x,{\mathbf k}_T^2,\mu^2)$ extended now into 
the ${\mathbf k}_T^2 > \mu^2$ region. This fact is in the clear contrast with the 
usual DGLAP evolution\footnote{We would like to note that 
cut-off $\Delta$ can be also taken $\Delta = |{\mathbf k}_T|/\mu$~[23]. In this case the unintegrated
parton distributions given by (1) --- (2) vanish for ${\mathbf k}_T^2 > \mu^2$ in accordance with 
the DGLAP strong ordering in ${\mathbf k}_T^2$.}.

The virtual (loop) contributions may be resummed 
to all orders by the quark and gluon Sudakov form factors
$$
  \ln T_q({\mathbf k}_T^2,\mu^2) = - \int\limits_{{\mathbf k}_T^2}^{\mu^2} {d {\mathbf p}_T^2\over {\mathbf p}_T^2} {\alpha_s({\mathbf p}_T^2)\over 2\pi} \int\limits_0^{z_{\rm max}} dz P_{qq}(z), \eqno (3)
$$
$$
  \ln T_g({\mathbf k}_T^2,\mu^2) = - \int\limits_{{\mathbf k}_T^2}^{\mu^2} {d {\mathbf p}_T^2\over {\mathbf p}_T^2} {\alpha_s({\mathbf p}_T^2)\over 2\pi} \left[ n_f \int\limits_0^1 dz P_{qg}(z) + \int\limits_{z_{\rm min}}^{z_{\rm max}} dz z P_{gg}(z) \right], \eqno (4)
$$

\noindent
where $z_{\rm max} = 1 - z_{\rm min} = {\mu/({\mu + |{\mathbf p}_T|}})$.
The form factors $T_a({\mathbf k}_T^2,\mu^2)$ give the probability of 
evolving from a scale ${\mathbf k}_T^2$ to a scale $\mu^2$ without 
parton emission. According to (3) and (4) $T_a({\mathbf k}_T^2,\mu^2) = 1$ 
in the ${\mathbf k}_T^2 > \mu^2$ region.

We would like to note that such definition of the $f_a(x,{\mathbf k}_T^2,\mu^2)$ is 
correct for ${\mathbf k}_T^2 > \mu_0^2$ only, where 
$\mu_0 \sim 1$ GeV is the minimum scale for which DGLAP evolution of the collinear
parton densities is valid. Everywhere in our numerical calculations 
we set the starting scale $\mu_0$ to be equal $\mu_0 = 1$ GeV.
Since the starting point of our derivation is the leading order 
DGLAP equations, the unintegrated parton distributions must satisfy
the normalisation condition
$$
  a(x,\mu^2) = \int\limits_0^{\mu^2} f_a(x,{\mathbf k}_T^2,\mu^2) d{\mathbf k}_T^2. \eqno(5)
$$

\noindent
This relation will be exactly satisfied if we define~[20]
$$
  f_a(x,{\mathbf k}_T^2,\mu^2)\vert_{{\mathbf k}_T^2 < \mu_0^2} = a(x,\mu_0^2) T_a(\mu_0^2,\mu^2). \eqno(6)
$$

\noindent
Then, we have obtained the unintegrated parton distributions $f_a(x,{\mathbf k}_T^2,\mu^2)$
in a proton. In order to obtain the unintegrated parton distribution $f_a^\gamma(x,{\mathbf k}_T^2,\mu^2)$ 
in a photon the same formulas (1) --- (4) can be also used~[21]. In the last case the
conventional parton distributions $a(x,\mu^2)$ in a proton should be replaced by
the corresponding parton densities $a^\gamma(x,\mu^2)$ in a photon.

In Figure~1 we show unintegrated parton densities $f_a(x,{\mathbf k}_T^2,\mu^2)$ in a proton at 
scale $\mu^2 = 100\,{\rm GeV}^2$ as a function 
of $x$ for different values of ${\mathbf k}_T^2$, namely 
${\mathbf k}_T^2 = 2\,{\rm GeV}^2$ (a),
${\mathbf k}_T^2 = 5\,{\rm GeV}^2$ (b),
${\mathbf k}_T^2 = 10\,{\rm GeV}^2$ (c) and 
${\mathbf k}_T^2 = 20\,{\rm GeV}^2$ (d). 
The solid, dashed, short dashed, dotted and dash-dotted lines
correspond to the unintegrated $u + \bar u$, $d + \bar d$, $s$, $c$ and gluon 
(divided by factor $10$) distributions, respectively. We have used here the standard 
GRV (LO) parametrizations~[27] of the collinear quark and gluon densities $a(x,\mu^2)$.
In order to be sure that normalization condition (5) is correctly 
satisfied we have performed the numerical integration of the parton densities 
$f_a(x,{\mathbf k}_T^2,\mu^2)$ over transverse momenta ${\mathbf k}_T^2$.
So, in Figure~2 we show our result for effective $u + \bar u$ quark and 
gluon (also divided by factor 10) distributions  
for different scales 
$\mu^2 = 2\,{\rm GeV}^2$ (a),
$\mu^2 = 5\,{\rm GeV}^2$ (b),
$\mu^2 = 10\,{\rm GeV}^2$ (c),
$\mu^2 = 20\,{\rm GeV}^2$ (d).
The solid lines correspond to the effective 
densities obtained from the unintegrated 
ones using relation (5). The dashed lines correspond to the collinear GRV (LO) parton 
distributions. One can see that normalization
condition (5) is exactly satisfied practically for all $x$ and $\mu^2$ values.
There are only rather small (less then few percent) 
violations of (5) in the case of the $u + \bar u$ quark distributions 
at large $x > 0.2$. We have checked numerically that the expression (5) is true also for  
other parton densities in a proton and in a photon.

For comparison we plot in Figure~2 (as dash-dotted lines) 
the corresponding LO parton distributions obtained by the CTEQ collaboration~[28] 
(CTEQ5L set). It is clear that there are some differences in both 
normalization and shape between GRV and CTEQ parametrizations.
In general, the CTEQ curves lie below the GRV ones by about $10$\%. 
This difference tends to be small when scale $\mu^2$ is large. 
However, the CTEQ collaboration does not provide a set of the 
parton distributions in a photon, which are necessary for 
calculation of the resolved contributions.
Therefore everywhere in our numerical analysis
we will use only the GRV parametrizations.

It is interesting to compare the KMR-constructed unintegrated parton densities  
with the distributions obtained in other approaches. Recently the full 
CCFM equation in a proton and in a photon was solved numerically using a Monte Carlo 
method, and new fits of the unintegrated gluon 
distributions (J2003 set 1 --- 3) have been presented~[29].
The input parameters were fitted to describe the proton structure function $F_2(x,Q^2)$.
These unintegrated gluon densities were used also in description of 
the forward jet production 
at HERA, charm and bottom production at Tevatron~[29], and charm and $J/\psi$ production at 
LEP2 energies~[30]. In Figure 3 we plot KMR (as solid lines) and J2003 set 1 (as dashed lines) 
gluon distributions in a proton
at scale $\mu^2 = 100\,{\rm GeV}^2$ as a function 
of $x$ for different values of ${\mathbf k}_T^2$, namely 
${\mathbf k}_T^2 = 2\,{\rm GeV}^2$ (a),
${\mathbf k}_T^2 = 10\,{\rm GeV}^2$ (b),
${\mathbf k}_T^2 = 20\,{\rm GeV}^2$ (c) and 
${\mathbf k}_T^2 = 50\,{\rm GeV}^2$ (d). 
One can see that J2003 set 1 gluon density is less steep at small $x$ compared to the
KMR one. The KMR gluon lie below J2003 set 1 at small ${\mathbf k}_T^2$ region
for $x > 3 \cdot 10^{-3}$. Typically the difference between solid and dashed lines
is about 30\% --- 40\% at $x = 0.01$.
This fact results to the some underestimation of the calculated 
cross sections in the KMR approach. This 
underestimation is about 30\% at HERA and 50\% at Tevatron\footnote{See Ref.~[13] for more details.}.
Therefore we can expect a rather large sensitivity of our predictions to the parton evolution 
scheme. 

We would like to point out again that behaviour of different unintegrated
parton distributions in a proton 
in the small ${\mathbf k}_T^2$ region (which essentially drives
the total cross sections) is very different, as it is clear shown on Figure~3.
However, the CCFM evolution does not include quark-initiated chains and therefore 
can not be used in our analysis since prompt photon production at HERA 
strongly depends on the quark distributions (see Section~3). Therefore in
our following investigations we will use only the KMR unintegrated parton densities.
But one should remember that dependence of our results on the evolution scheme 
may be rather large, and further theoretical attempts (in order to investigate 
the unintegrated quark distributions more detail) are necessary to reduce this uncertainty.

\section{Calculation details} 

\subsection{The subprocesses under consideration} \indent 

In $ep$ collisions at HERA prompt photons can be produced by one 
of three mechanisms: a direct production, a single resolved production 
and via parton-to-photon fragmentation processes~[5].
The direct contribution to the $\gamma p \to \gamma + X$ process
is the Deep Inelastic Compton (DIC) scattering on the quark (antiquark)
$$
  \gamma (k_1) + q (k_2) \to \gamma (p_{\gamma}) + q (p'), \eqno(7)
$$

\noindent
where the particles four-momenta are given in parentheses. It gives 
the ${\cal O}(\alpha^2_{em})$ order contribution to the hadronic
cross section. Here $\alpha_{em}$ is Sommerfeld's fine structure constant.
The single resolved processes are
$$
  q (k_1) + g (k_2) \to \gamma (p_{\gamma}) + q (p'), \eqno(8)
$$
$$
  g (k_1) + q (k_2) \to \gamma (p_{\gamma}) + q (p'), \eqno(9)
$$
$$
  q (k_1) + q (k_2) \to \gamma (p_{\gamma}) + g (p'). \eqno(10)
$$

\noindent
Since the parton distributions in a photon $a_{\gamma}(x,\mu^2)$ 
have a leading behavior proportional to $\alpha_{em}\ln \mu^2/\Lambda_{\rm QCD}^2 \sim \alpha_{em}/\alpha_s$,
these subprocesses give also
the ${\cal O}(\alpha^2_{em})$ contributions and 
therefore should be taken into account in our analysis.

In addition to the direct and resolved production, photons can
be also produced through the fragmentation of a hadronic jet
into a single photon carrying a large fraction $z$ of the jet energy~[5].
These processes are described in terms of quark-to-photon $D_{q\to\gamma}(z,\mu^2)$ and
gluon-to-photon $D_{g\to\gamma}(z,\mu^2)$ fragmentation functions~[7]. 
The main feature of the fragmentation contribution in leading order is fact that
produced photon balanced by a jet on the opposite side of the event
and accompanied by collinear hadrons on the same side of the event.

It is very important that in order to reduce the huge background
from the secondary photons produced by the decays of $\pi^0$, $\eta$ 
and $\omega$ mesons the isolation criterion is introduced in the experimental analyses.
This criterion is the following. A photon is isolated if the 
amount of hadronic transverse energy $E_T^{\rm had}$, deposited inside
a cone with aperture $R$ centered around the photon direction in the 
pseudo-rapidity and azimuthal angle plane, is smaller than
some value $E_T^{\rm max}$:
$$
  \displaystyle E_T^{\rm had} \le E_T^{\rm max},\atop
  \displaystyle (\eta - \eta^{\gamma})^2 + (\phi - \phi^{\gamma})^2 \le R^2. \eqno(11)
$$

\noindent
The both H1 and ZEUS collaborations take $R = 1$, $E_T^{\rm max} = \epsilon E_T^\gamma$ with 
$\epsilon = 0.1$ in the experiment~[1--4]. 
Isolation not only reduces the background 
but also significantly reduces the fragmentation components.
It was shown~[7] that after applying the isolation cut (11) the contribution from the
fragmentation subprocesses is about 5 --- 6\% of the total cross section.
Since the dependence of our results on the non-collinear parton evolution scheme may be 
rather large (as it was demonstrated in Section~2), in our further analysis we will 
neglect the relative small fragmentation contribution and consider only the 
direct and resolved production 
(7) --- (10). We note that photon produced in these processes is automatically 
isolated from the quark or gluon jet by requiring a non-zero transverse momentum
of a photon or jet in the $\gamma p$ center-of-mass frame.

It was claimed~[7, 25] that direct box diagram $\gamma g\to \gamma g$, which is
formally of the next-to-next-to-leading order (NNLO), gives approximately 6\% contribution 
to the total NLO cross section. In the present paper we will not take into account 
this diagram also.

\subsection{Cross section for prompt photon production} \indent 

Let $p_e$ and $p_p$ be the four-momenta of the initial electron and proton.
The direct contribution (7) to the $\gamma p \to \gamma + X$ 
process in the $k_T$-factorization approach can be written as
$$
  d\sigma^{\rm (dir)} (\gamma p \to \gamma + X) = \sum_{q} \int {dx_2\over x_2} f_q(x_2,{\mathbf k}_{2 T}^2,\mu^2) d{\mathbf k}_{T}^2 {d\phi_2\over 2\pi} d{\hat \sigma} (\gamma q \to \gamma q), \eqno(12)
$$

\noindent 
where ${\hat \sigma} (\gamma q \to \gamma q)$ is the hard subprocess cross section 
via quark or antiquark having fraction $x_2$ of a initial proton longitudinal 
momentum, non-zero transverse momentum ${\mathbf k}_{2 T}$ 
(${\mathbf k}_{2 T}^2 = - k_{2 T}^2 \neq 0$) and azimuthal angle $\phi_2$.
The expression (12) can be easily rewritten in the form
$$
  \displaystyle \sigma^{\rm (dir)} (\gamma p \to \gamma + X) = \sum_q \int {E_T^\gamma \over 8\pi (x_2 s)^2 (1 - \alpha)} |\bar {\cal M}|^2 (\gamma q \to \gamma q) \times \atop 
  \displaystyle \times f_q(x_2,{\mathbf k}_{2 T}^2,\mu^2) dy^\gamma d E_T^\gamma d{\mathbf k}_{2 T}^2 {d\phi_2\over 2\pi} {d\phi^\gamma\over 2\pi}, \eqno(13)
$$

\noindent
where $|\bar {\cal M}|^2 (\gamma q \to \gamma q)$ is the hard
matrix element which depends on the transverse momentum ${\mathbf k}_{2 T}^2$, 
$s = (k_1 + p_p)^2$ is the total energy of the subprocess under consideration, 
$y^\gamma$, $E_T^\gamma$ 
and $\phi^\gamma$ are the rapidity, transverse energy and azimuthal angle of the 
produced photon in the $\gamma p$ center-of-mass frame, and $\alpha = E_T^\gamma \exp y^\gamma /\sqrt s$. 

The formula for the 
resolved contribution to the prompt photon photoproduction in the 
$k_T$-factorization approach can be obtained by the similar way. But one should keep in mind that
convolution in (12) should be made also with the unintegrated parton distributions 
$f_a^\gamma(x,{\mathbf k}_{T}^2,\mu^2)$ in a photon, i.e.  
$$
  \displaystyle d\sigma^{\rm (res)} (\gamma p \to \gamma + X) = \sum_{a,b} \int {dx_1\over x_1} f_a^\gamma(x_1,{\mathbf k}_{1 T}^2,\mu^2) d{\mathbf k}_{1 T}^2 {d\phi_1\over 2\pi} \times \atop 
  \displaystyle \times \int {dx_2\over x_2} f_b(x_2,{\mathbf k}_{2 T}^2,\mu^2) d{\mathbf k}_{2 T}^2 {d\phi_2\over 2\pi} d{\hat \sigma} (a b \to \gamma c), \eqno(14)
$$

\noindent
where $a, b, c = q$ and/or $g$, ${\hat \sigma} (a b \to \gamma c)$ is the cross section 
of the photon production in the corresponding parton-parton interaction (8) --- (10). 
Here parton $a$ has fraction $x_1$ of a initial photon longitudinal 
momentum, non-zero transverse momentum ${\mathbf k}_{1 T}$ 
(${\mathbf k}_{1 T}^2 = - k_{1 T}^2 \neq 0$) and azimuthal angle $\phi_1$.
We can easily obtain the final expression from equation (14). It has the form
$$
  \displaystyle \sigma^{\rm (res)} (\gamma p \to \gamma + X) = \sum_{a,b} \int {E_T^\gamma\over 8\pi (x_1 x_2 s)^2} |\bar {\cal M}|^2(ab \to \gamma c) \times \atop 
  \displaystyle \times f_a^\gamma(x_1,{\mathbf k}_{1 T}^2,\mu^2) f_b(x_2,{\mathbf k}_{2 T}^2,\mu^2) d{\mathbf k}_{1 T}^2 d{\mathbf k}_{2 T}^2 dE_T^\gamma dy^\gamma dy^c {d\phi_1\over 2\pi} {d\phi_2\over 2\pi} {d\phi^\gamma\over 2\pi}, \eqno(15)
$$

\noindent
where $y^c$ is the rapidity of the parton $c$ in the $\gamma p$ center-of-mass frame. It is 
important that hard matrix elements $|\bar {\cal M}|^2 (ab \to \gamma c)$ depend
on the transverse momenta ${\mathbf k}_{1 T}^2$ and ${\mathbf k}_{2 T}^2$.
We would like to note that if we average the expressions (13) and (15) over 
${\mathbf k}_{1 T}$ and ${\mathbf k}_{2 T}$ and take 
the limit ${\mathbf k}_{1 T}^2 \to 0$ and ${\mathbf k}_{2 T}^2 \to 0$,
then we obtain well-known expressions for 
the prompt photon production in leading-order (LO) perturbative QCD.

The experimental data taken by the H1~[4] and ZEUS~[3] collaborations refer to the prompt
photon production in the $ep$ collisions, where electron 
is scattered at small angle and the mediating photon is almost real ($Q^2 \sim 0$).
Therefore $\gamma p$ cross sections (13) and (15)
needs to be weighted with the photon flux in the electron:
$$
  d\sigma(e p\to e' + \gamma + X) = \int f_{\gamma/e}(y) d\sigma(\gamma p \to \gamma + X) dy, \eqno (16)
$$

\noindent
where $y$ is a fraction of the initial electron energy taken by the photon in the laboratory frame,
and we use the Weizacker-Williams approximation for the bremsstrahlung photon
distribution from an electron:
$$
  f_{\gamma/e}(y) = {\alpha_{em} \over 2\pi}\left({1 + (1 - y)^2\over y}\ln{Q^2_{\rm max}\over Q^2_{\rm min}} + 
  2m_e^2 y\left({1\over Q^2_{\rm max}} - {1\over Q^2_{\rm min}} \right)\right). \eqno (17)
$$

\noindent
Here $m_e$ is the electron mass, 
$Q^2_{\rm min} = m_e^2y^2/(1 - y)^2$ and $Q^2_{\rm max} = 1\,{\rm GeV}^2$, 
which is a typical value for the recent photoproduction measurements at the HERA collider.

The multidimensional integration in (13), (15) and (16) has been performed
by means of the Monte Carlo technique, using the routine VEGAS~[31].
The full C$++$ code is available from the authors on 
request\footnote{lipatov@theory.sinp.msu.ru}.
For reader's convenience, we collect the analytic expressions for the 
off-shell matrix elements which correspond to all partonic subprocesses under consideration
(7) --- (10) in the Appendix. These formulas may be useful for the subsequent applications.

\section{Numerical results} \indent 

We now are in a position to present our numerical results. First we describe our
theoretical input and the kinematical conditions. After we fixed the unintegrated
parton distributions in a proton $f_a(x,{\mathbf k}_{T}^2,\mu^2)$ and in a photon 
$f_a^\gamma(x,{\mathbf k}_{T}^2,\mu^2)$, the cross sections (13) and (15) depend on
the energy scale $\mu$. As it often done~[7, 25] for prompt photon production, 
we choose the renormalization and
factorization scales to be $\mu = \xi E_T^\gamma$. 
In order to estimate the theoretical uncertainties of our calculations
we will vary the scale parameter $\xi$ between 1/2 and 2 about the default value $\xi = 1$.
Also we use LO formula
for the strong coupling constant $\alpha_s(\mu^2)$ with $n_f = 3$ active (massless) quark
flavours and $\Lambda_{\rm QCD} = 232$ MeV, such that $\alpha_s(M_Z^2) = 0.1169$.
In our analysis we will not neglect the charm quark mass and set it to be $m_c = 1.4$ GeV.

\subsection{Inclusive prompt photon production} \indent 

Experimental data for the inclusive prompt photon production $e p \to e' + \gamma + X$
comes from both ZEUS and H1 collaborations. Two differential cross section are determined: first
as a function of the transverse energy $E_T^\gamma$, and second as
a function of pseudo-rapidity $\eta^\gamma$. The ZEUS data~[3] refer to the
kinematic region\footnote{Here and in the following all kinematic quantities
are given in the laboratory frame where positive OZ axis direction is given by the proton beam.}
defined by $5 < E_T^\gamma < 10$ GeV and $-0.7 < \eta^\gamma < 0.9$ with
electron energy $E_e = 27.5$ GeV and proton energy $E_p = 820$ GeV.
The fraction $y$ of the electron energy trasferred to the photon is restricted
to the range $0.2 < y < 0.9$. Additionally the available ZEUS data for the 
prompt photon pseudo-rapidity distributions have been given also for three 
subdivisons of the $y$ range, namely $0.2 < y < 0.32$ ($134 < W < 170$ GeV), 
$0.32 < y < 0.5$ ($170 < W < 212$ GeV) and $0.5 < y < 0.9$ ($212 < W < 285$ GeV).
The more recent H1 data~[4] refer to the kinematic region
defined by $5 < E_T^\gamma < 10$ GeV, $ - 1 < \eta^\gamma < 0.9$ and $0.2 < y < 0.7$
with electron energy $E_e = 27.6$ GeV and proton energy $E_p = 920$ GeV.

The transverse energy distributions of the inclusive prompt photon for 
different kinematical region are shown
in Figures~4 and~5 in comparison to the HERA data. 
Instead of presenting our theoretical predictions as continuous lines, we adopt the
binning pattern encoded in the experimental data.
The solid histograms obtained by fixing both the
factorization and normalization scales at the default value $\mu = E_T^\gamma$,
whereas upper and lower dashed histograms correspond to the $\mu = E_T^\gamma/2$ and
$\mu = 2 E_T^\gamma$ scales, respectively. One can see that predicted
cross sections agree well with the experimental data
except the moderate $E_T^\gamma$ region.
We would like to note that overall agreement with data can be improved 
when unintegrated quark and gluon distributions in a proton and in a photon 
will be studied more detail. It is because the KMR approach tends to 
underestimate the calculated cross sections, as it was discussed in Section~2.
The collinear NLO QCD calculations~[7, 25] give the similar description of 
the transverse energy distributions measured by the ZEUS collaboration. 
At the same time, according to the analysis which was done by the H1 
collaboration~[4], in order to obtain a realistic comparison of 
their data and theory the corrections for hadronisation
and multiple interactions should be taken into account in the 
predictions\footnote{See Ref.~[4] for more details.}.
The correction factors are typically 0.7 --- 0.9 depending on a bin.
The NLO calculations~[7, 25] are approximately 30\% --- 40\% below the H1 data
if the corrections for hadronisation and multiple interactions are 
applied. We would like to note that these corrections are not accounted 
for in our analysis.
The effect of scale variations in transverse energy distributions 
is rather large: the relative difference
between results for $\mu = E_T^\gamma$ and results for $\mu = E_T^\gamma/2$ or
$\mu = 2 E_T^\gamma$ is about 15\%.

The pseudo-rapidity distributions of the inclusive prompt photon production
compared with the HERA data in different kinematical region are
shown in Figures~6 and~7. All histograms here are the same as in Figure~4.
One can see that measured distributions are reasonably
well described in the pseudo-rapidity region 
$ - 0.4 \leq \eta^\gamma \leq 0.9$ only. For $ - 1 \leq \eta^\gamma \leq -0.4$
our predictions lie mostly below the experimental 
points\footnote{Note that such disagreement between predicted and measured 
cross sections is observed for collinear NLO QCD calculations~[7, 25] also.}.
The discrepancy between data and theory at negative $\eta^\gamma$ is
found to be relative strong at low values of the initial photon
fractional momentum $y$. So, in Figures~8,~9 and~10 we show the 
inclusive cross sections $d\sigma/d\eta^\gamma$
evaluated for the three $y$ ranges $0.2 < y < 0.32$ ($134 < W < 170$ GeV), 
$0.32 < y < 0.5$ ($170 < W < 212$ GeV) and $0.5 < y < 0.9$ ($212 < W < 285$ GeV), 
respectively. All histograms here are the same as in Figure~4. In the lowest $y$ range, 
both our predictions and experimental data show a peaking at negative $\eta^\gamma$, 
but it is stronger in the data. In the high $y$ region, $0.5 < y < 0.9$, a good 
agreement is obtained. This fact allows to establish that the above
discussed discrepancy between the data and theory at $- 1 \leq \eta^\gamma \leq - 0.4$ is
coming from the low ($0.2 < y < 0.32$) and medium ($0.32 < y < 0.5$) $y$ region.
The scale variation changes the estimated 
cross sections by about 15\%. The collinear NLO QCD calculations~[7, 25] give the 
similar description of the pseudo-rapidity distributions 
measured by the ZEUS collaboration. At the same time, after 
corrections for hadronisation and multiple interactions (not accounted for in our analysis) 
the NLO predictions are 30\% --- 40\% below the H1 data.

As it was already mentioned above, the dependence of the our results on a 
renormalization/factorization scale $\mu$ is rather large, about 10\% --- 15\% in 
the wide kinematic range. 
There are also additional uncertainties come from the unintegrated parton densities,
as it was discussed in Section~2.
The theoretical uncertainties of the collinear NLO QCD calculations
are about 3\%~[7, 25]. This fact indicates that contribution from 
NNLO and high order terms is not significant. 
At the same time the strong scale dependence of our results demonstrates the necessarity 
of reducing of uncertainties in the non-collinear parton evolution.

The individual contributions from the direct and resolved production 
mechanisms to the total cross section in the $k_T$-factorization approach 
is about 47\% and 53\%, respectively. In the resolved conribution, the 
channels (8), (9) and (10) account for 80\%, 14\% and 6\%. 
Additionally, using the Duke-Owens (DO)~[32] parton-to-photon fragmentation functions,
we perform the estimation of the fragmentation component (not shown in Figures).
We find that after applying isolation cut it give only a very small (about few percent) 
contribution.

\subsection{Prompt photon production in association with jet} \indent 

Now we demonstrate how $k_T$-factorization approach can be 
used to calculate the semi-inclusive prompt photon production rates.
The produced photon is
accompanied by a number of partons radiated in the course of the parton evolution.
As it has been noted in Ref.~[33], on the average the parton 
transverse momentum decreases from the hard interaction
block towards the proton. As an approximation, we assume that the parton $k'$ 
emitted in the last evolution step compensates the whole transverse momentum
of the parton participating in the hard subprocess, i.e. ${\mathbf k'}_{T} \simeq - {\mathbf k}_{T}$.
All the other emitted partons are collected together in the proton remnant, which
is assumed to carry only a negligible transverse momentum compared to ${\mathbf k'}_{T}$.
This parton gives rise to a final hadron jet with $E_T^{\rm jet} = |{\mathbf k'}_{T}|$
in addition to the jet produced in the hard subprocess. From these hadron jets
we choose the one carrying the largest transverse energy, and then compute prompt
photon with an associated jet cross sections.

Experimental data for such processes were obtained~[4] very recently by the H1 
collaboration. The cross sections measured differentially as a function of 
$E_T^\gamma$, $E_T^{\rm jet}$,
and the pseudo-rapidities $\eta^\gamma$ and $\eta^{\rm jet}$ in the 
kinematic region defined by $5 < E_T^\gamma < 10$ GeV, $E_T^{\rm jet} > 4.5$ GeV, 
$ - 1 < \eta^\gamma < 0.9$, $ - 1 < \eta^{\rm jet} < 2.3$ and $0.2 < y < 0.7$ with
electron energy $E_e = 27.6$ GeV and proton energy $E_p = 920$ GeV.
There are no ZEUS data for the prompt photon plus jet production, 
although some data for distribution of events, not corrected for the detector 
effects, were presented~[2].

The transverse energy $E_T^\gamma$ and pseudo-rapidity $\eta^\gamma$ distributions 
of the prompt photon plus jet production are shown in Figures~11 and~12 in 
comparison with H1 data.
All histograms here are the same as in Figure~4. In contrast to the inclusive case, one 
can see that our predictions are consistent with the data in most bins, although some
discrepancies are present. The scale variation as it was described above 
changes the estimated cross sections by about 10\%. The results of the 
collinear NLO calculations~[7, 25] 
which include corrections for hadronisation and multiple interactions give the 
similar results and consistent with data also.

In Figures~13 and~14 we show our predictions for the transverse energy $E_T^{\rm jet}$ and 
pseudo-rapidity $\eta^{\rm jet}$ distributions in comparison with H1 data.
All histograms here are the same as in Figure~4. A rather good agreement between
our results and data is obtained again. It is interesting to note that 
shape of the predicted pseudo-rapidity $\eta^{\rm jet}$ distribution coincide 
with the one obtained in the collinear NLO calculations~[7, 25].
At the same time the shape of this distribution 
is not reproduced by the leading-order QCD calculations~[4].
This fact can demonstrate again that the main part of the collinear 
high-order corrections is already included at LO level in 
$k_T$-factorization formalism.
The scale dependence of our predictions is about 10\%.

The most important variables for testing the structure of colliding 
proton and photon are the longitudinal fractional momenta of partons in these particles.
In order to reconstruct the momentum fractions of the 
initial partons from measured quantities the observables $x_\gamma$ and
$x_p$ are introduced~[4]:
$$
  \displaystyle x_\gamma = {E_T^\gamma ( e^{ - \eta^\gamma} + e^{ - \eta^{\rm jet}} )\over 2 y E_e}, \quad 
  \displaystyle x_p = {E_T^\gamma ( e^{\eta^\gamma} + e^{\eta^{\rm jet}} )\over 2 E_p}. \eqno(18)
$$

\noindent
These observables make explicit use only of the photon energy, which is better measured
than the jet energy. The $x_\gamma$ distribution is particularly sensitive to the
photon structure function. At large $x_\gamma$ region ($x_\gamma > 0.85$) the cross
section is dominated by the contribution of processes with direct initial photons, whereas
at $x_\gamma < 0.85$ the resolved photon contributions dominate~[4].

So, in Figures~15 and~16 the $x_\gamma$ and $x_p$ distributions are shown in comparison with H1 data.
One can see that our 
predictions reasonable agree with experimental data. The NLO calculations~[7, 25] without corrections
for hadronisation and multiple interactions give the similar results. However, NLO calculations
tend to underestimate the H1 data if these corrections are taken into account. The hadronic and
multiple interaction corrections improve the description of the data at $x_\gamma < 0.6$ only~[4].

Further understanding of the process dynamics and in particular of the high-order correction effects 
may be obtained from the transverse correlation between the produced prompt photon and the jet.
The H1 collaboration has measured the distribution on the component of the
prompt photon's momentum perpendicular to the jet direction in the transverse plane, i.e.
$$
  p_T = |{\mathbf p}_{T}^\gamma \times {\mathbf p}_{T}^{\rm jet}|/|{\mathbf p}_{T}^{\rm jet}| = E_T^\gamma \sin \Delta \phi, \eqno(19)
$$

\noindent
where $\Delta \phi$ is the difference in azimuth between the photon and the jet. In the
collinear leading order approximation, the distribution over $p_T$ must be simply
a delta function $\delta(p_T)$, since the produced photon and the jet are back-to-back
in the transverse plane. Taking into account the non-vanishing initial parton
transverse momenta ${\mathbf k}_{1 T}$ and ${\mathbf k}_{2 T}$ leads to the violation
of this back-to-back kinematics in the $k_T$-factorization approach.

The normalised $p_T$ distributions are shown in Figures~17 and~18 separately for the 
regions $x_\gamma < 0.85$ and $x_\gamma > 0.85$, where direct and resolved photon induced
processes dominate, respectively. All histograms here are the same as in Figure~4.
Our predictions are consistent with the H1 data for all $x_\gamma$ values 
except large $p_T$ region. So, at $p_T > 5$ GeV the results of our calculations lie slightly below the 
data at $x_\gamma < 0.85$ and above the data at $x_\gamma > 0.85$.
At the same time the NLO QCD prediction~[7] gives a better description of the $p_T$ 
distributions at $x_\gamma < 0.85$ than another one~[25]. It is because in this region
the cross section is dominated by ${\cal O}(\alpha_s)$ corrections to the processes with
resolved photons, which are not included in the NLO calculations~[25]. 
In general, we can conclude that our results lie between the predictions~[7] and the predictions~[25]
in the whole $x_\gamma$ range. This fact indicates again that the 
main part of the high-order collinear 
corrections is effectively included in our calculations.

Finally, we would like to note that there are, of course, still rather 
large theoretical uncertainties in our results connected with 
unintegrated parton distributions, and it is necessary to work hard until 
these uncertainties will be reduced. However, it was shown~[24] that the
properties of different unintegrated parton distributions clear manifest themselves in the 
azimuthal correlation between transverse momenta of the final state particles.
Therefore we can expect that further theoretical and experimental studying of 
these correlations will give important information about non-collinear parton evolution
dynamics in a proton and in a photon.

\section{Conclusions} \indent 

We have investigated the prompt photon photoproduction at the HERA collider in the
$k_T$-factorization approach. In order to obtain the unintegrated quark and gluon distributions 
in a proton and in a photon we used the Kimber-Martin-Ryskin prescription. We have investigated
both inclusive and associated with jet prompt photon production rates. Such calculations 
in the $k_T$-factorization approach were performed for the first time.

We took into account both the direct and resolved contributions and investigated
the sensitivity of the our results to renormalization and factorization scales. There are,
of course, also theoretical uncertainties due to non-collinear evolution scheme.
However, much more work needs to be done before these uncertainties will be reduced.

We have found that our predictions for the inclusive prompt photon production 
are in reasonable agreement with the H1 and ZEUS data except rear (electron direction) pseudo-rapidity
region. In contrast, our results for prompt photon associated with jet are
consistent with data in the whole kinematical range. However, the scale dependence of our results
is rather large compared to the collinear NLO QCD calculations.
At the same time we demonstrate that main part of the standard 
high-order corrections is already included in the $k_T$-factorization formalism at
LO level.

Note that in our analysis we neglect the contribution from the fragmentation processes
and from the direct box diagram ($\gamma g \to \gamma g$). Since the relative large
box contribution (about 6\% of the total NLO cross section) is mainly due to large gluonic 
content of the proton at small $x$, studying of this subprocess should be also
very interesting in the $k_T$-factorization approach. 
We plan to investigate it in detail in the forthcoming publications.

\section{Acknowledgements} \indent 

The authors are very grateful also to S.P.~Baranov for encouraging interest
and helpful discussions. This research was supported in part by the FASI of Russian Federation.

\section{Appendix} \indent 

Here we present the compact analytic expressions for the 
hard matrix elements which appear in (13) and (15). In the 
following, $\hat s$, $\hat t$ and $\hat u$ are usual Mandelstam 
variables for corresponding $2 \to 2$ subprocesses and
$e_q$ is the fractional electric charge of quark $q$.

We start from the direct subprocess (7). The corresponding
squared matrix element summed over final polarization states and 
averaged over initial ones read
$$
  |\bar {\cal M}|^2(\gamma q \to \gamma q) = {\displaystyle 2 (4\pi)^2 \alpha_{em}^2 e_q^4 \over 
  \displaystyle (\hat s - m^2)^2 (\hat u - m^2)^2 } F_{\gamma q} ({\mathbf k}_{2 T}^2), \eqno (A.1)
$$

\noindent
where $m$ is the quark mass, and
$$
  F_{\gamma q} ({\mathbf k}_{T}^2) = 6 m^8 - (3 \hat s^2 + 14 \hat s \hat u + 3 \hat u^2) m^4 + (\hat s^3 + 7 \hat s^2 \hat u + \atop 7 \hat u^2 \hat s + \hat u^3) m^2 - (\hat s^2 + \hat u^2) \hat s \hat u. \eqno(A.2) 
$$

\noindent
It is important to note that when we calculate the Dirac's traces we set the incoming 
quark four-momentum to be equal $k_2 = x_2 p_p$. Therefore these formulas formally are the same as in 
the usual leading-order collinear approach and there is no obvious dependence on the 
parton transverse momentum ${\mathbf k}_{2 T}$. However, this dependence is present because 
we have used true off-shell kinematics in order to 
estimate the cross section (13).
It is in the clear contrast with the collinear calculations.

The squared matrix elements of the resolved photon contributions (8) --- (10) 
summed over final polarization states and averaged over initial ones read
$$
  |\bar {\cal M}|^2(q g \to \gamma q) = {\displaystyle (4\pi)^2 \alpha_{em} \alpha_s e_q^2 \over 
  \displaystyle 3 (\hat t - m^2)^2 (\hat s - m^2)^2 } F_{q g} ({\mathbf k}_{1 T}^2, {\mathbf k}_{2 T}^2), \eqno (A.3)
$$
$$
  |\bar {\cal M}|^2(g q \to \gamma q) = {\displaystyle (4\pi)^2 \alpha_{em} \alpha_s e_q^2 \over 
  \displaystyle 3 (\hat s - m^2)^2 (\hat u - m^2)^2 } F_{g q} ({\mathbf k}_{1 T}^2, {\mathbf k}_{2 T}^2), \eqno (A.4)
$$
$$
  |\bar {\cal M}|^2(q q \to \gamma g) = - {\displaystyle 8 (4\pi)^2 \alpha_{em} \alpha_s e_q^2 \over 
  \displaystyle 9 (\hat t - m^2)^2 (\hat u - m^2)^2 } F_{q q} ({\mathbf k}_{1 T}^2, {\mathbf k}_{2 T}^2), \eqno (A.5)
$$

\noindent where functions $F_{qg}({\mathbf k}_{1 T}^2, {\mathbf k}_{2 T}^2)$, 
$F_{gq}({\mathbf k}_{1 T}^2, {\mathbf k}_{2 T}^2)$ and $F_{qq}({\mathbf k}_{1 T}^2, {\mathbf k}_{2 T}^2)$ are given by
$$
  F_{q g} ({\mathbf k}_{1 T}^2, {\mathbf k}_{2 T}^2) = 6 m^ 8 - (2 {\mathbf k}_{2 T}^4 + 2 (\hat s + \hat t) {\mathbf k}_{2 T}^2 + 3 \hat s^2 + 3 \hat t^2 + 14 \hat s \hat t)m^4 + 
$$
$$
  (2 (\hat s + \hat t) {\mathbf k}_{2 T}^4 + 8 \hat s \hat t {\mathbf k}_{2 T}^2 + \hat s^3 + \hat t^3 + 7 \hat s \hat t^2 + 7 \hat s^2 \hat t) m^2 - 
$$
$$
   \hat s \hat t (2 {\mathbf k}_{2 T}^4 + 2 (\hat s + \hat t) {\mathbf k}_{2 T}^2 + \hat s^2 + \hat t^2), \eqno(A.6)
$$
$$
  F_{g q} ({\mathbf k}_{1 T}^2, {\mathbf k}_{2 T}^2) = 6 m^ 8 - (2 {\mathbf k}_{1 T}^4 + 2 (\hat s + \hat u) {\mathbf k}_{1 T}^2 + 3 \hat s^2 + 3 \hat u^2 + 14 \hat s \hat u)m^4 + 
$$
$$
  (2 (\hat s + \hat u) {\mathbf k}_{1 T}^4 + 8 \hat s \hat u {\mathbf k}_{1 T}^2 + \hat s^3 + \hat u^3 + 7 \hat s \hat u^2 + 7 \hat s^2 \hat u) m^2 - 
$$
$$
   \hat s \hat u (2 {\mathbf k}_{1 T}^4 + 2 (\hat s + \hat u) {\mathbf k}_{1 T}^2 + \hat s^2 + \hat u^2), \eqno(A.7)
$$
$$
  F_{q q} ({\mathbf k}_{1 T}^2, {\mathbf k}_{2 T}^2) = 6 m^8 - (3 \hat t^2 + 3 \hat u^2 + 14 \hat t \hat u) m^4 + (\hat t^3 + \hat u^3 + \atop 
  7 \hat t \hat u^2 + 7 \hat t^2 \hat u) m^2 - \hat t \hat u (\hat t^2 + \hat u^2). \eqno(A.8)
$$

\noindent
Since we take into account the ${\mathbf k}_{T}$ depencence of the incoming virtual gluon polarization 
tensor, the functions $F_{q g} ({\mathbf k}_{1 T}^2, {\mathbf k}_{2 T}^2)$ and 
$F_{g q} ({\mathbf k}_{1 T}^2, {\mathbf k}_{2 T}^2)$ also depend obviously on the 
gluon transverse momentum. It is clear that if we take 
the limit ${\mathbf k}_{1 T}^2 \to 0$, ${\mathbf k}_{2 T}^2 \to 0$ in (A.1) --- (A.8) we 
easily obtain the corresponding collinear formulas.

Finally, we would like to point out again that in numerical computations we use precise 
off-shell kinematics and therefore all expressions (A.1) --- (A.8) depends 
on the parton transverse momentum. In particular, the incident 
parton momentum fractions $x_1$ and $x_2$ in (13) and (15) have some ${\mathbf k}_{T}$ dependence.
In the limit ${\mathbf k}_{1 T} \to 0$, ${\mathbf k}_{2 T} \to 0$ we reproduce 
standard leading-order QCD collinear results.

\newpage

\begin{figure}
\epsfig{figure=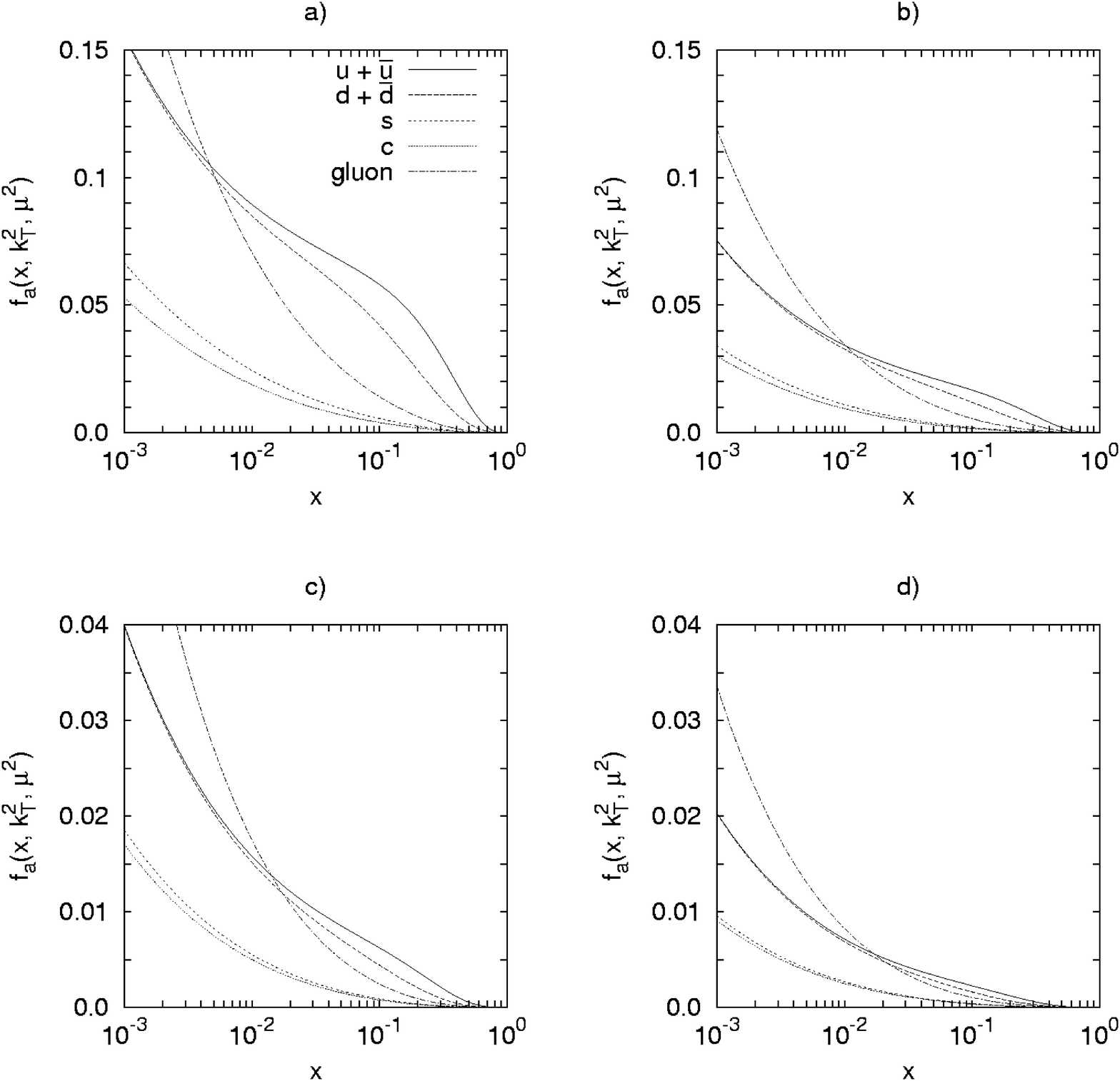, width = 17.5cm, height = 17.5cm}
\caption{The unintegrated parton distributions $f_a(x,{\mathbf k}_T^2,\mu^2)$
at scale $\mu^2 = 100\,{\rm GeV}^2$ as a function 
of $x$ for different values of ${\mathbf k}_T^2$, namely 
${\mathbf k}_T^2 = 2\,{\rm GeV}^2$ (a),
${\mathbf k}_T^2 = 5\,{\rm GeV}^2$ (b),
${\mathbf k}_T^2 = 10\,{\rm GeV}^2$ (c) and 
${\mathbf k}_T^2 = 20\,{\rm GeV}^2$ (d).
The solid, dashed, short dashed, dotted and dash-dotted lines
correspond to the unintegrated $u + \bar u$, $d + \bar d$, $s$, $c$ and gluon 
(divided by factor $10$) distributions, respectively.}
\label{fig1}
\end{figure}

\newpage

\begin{figure}
\epsfig{figure=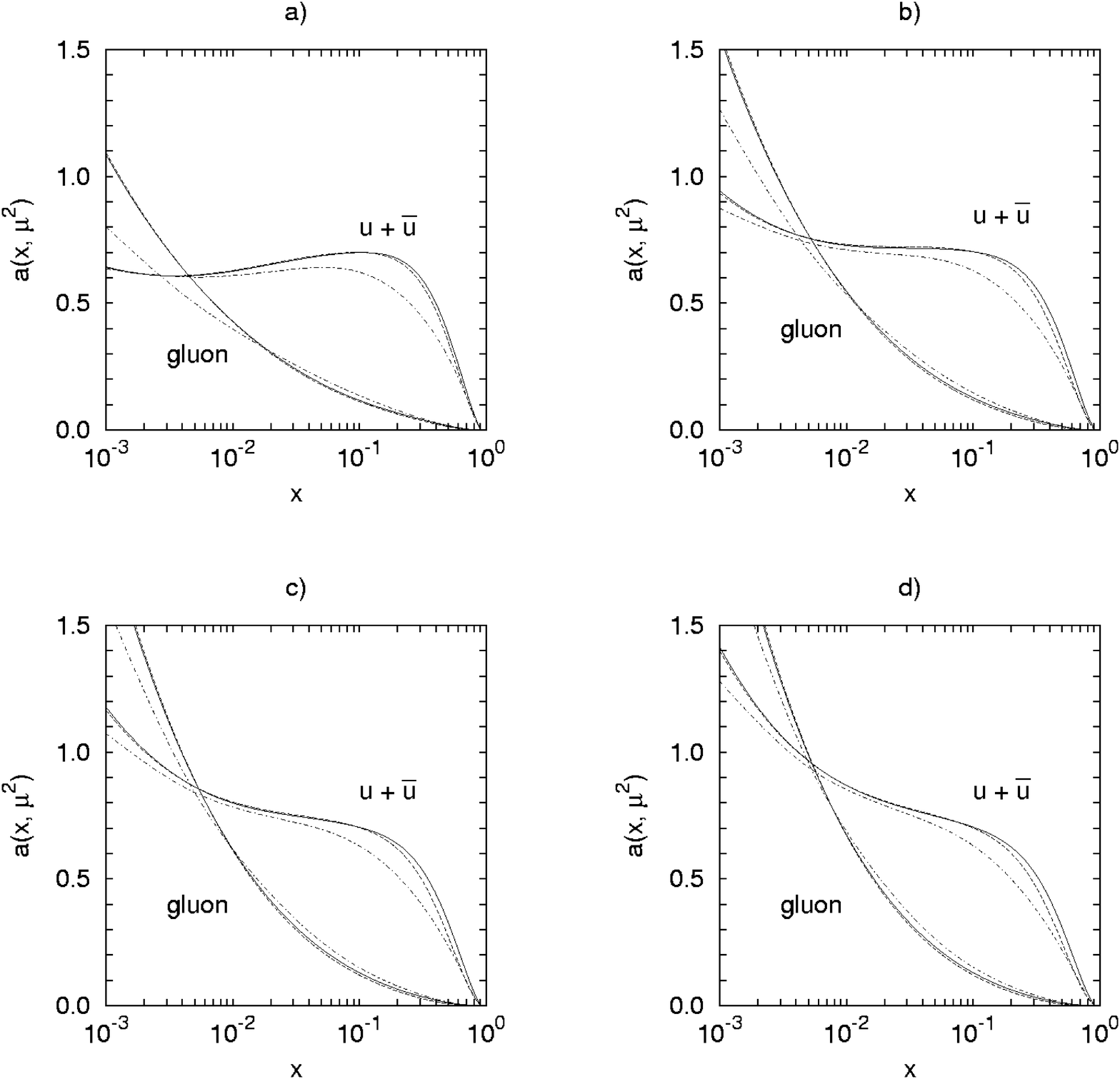, width = 17.5cm, height = 17.5cm}
\caption{The effective parton distributions $a(x,\mu^2)$
as a function of $x$ for different values of $\mu^2$, namely 
$\mu^2 = 2\,{\rm GeV}^2$ (a),
$\mu^2 = 5\,{\rm GeV}^2$ (b),
$\mu^2 = 10\,{\rm GeV}^2$ (c) and 
$\mu^2 = 20\,{\rm GeV}^2$ (d).
The solid lines correspond to the parton densities taken from the 
unintegrated ones using relation (5). The dashed and dash-dotted lines 
correspond to the conventional GRV (LO) and CTEQ5L parton distributions, respectively.
Everywhere the gluon distributions are divided by factor $10$.}
\label{fig2}
\end{figure}

\newpage

\begin{figure}
\epsfig{figure=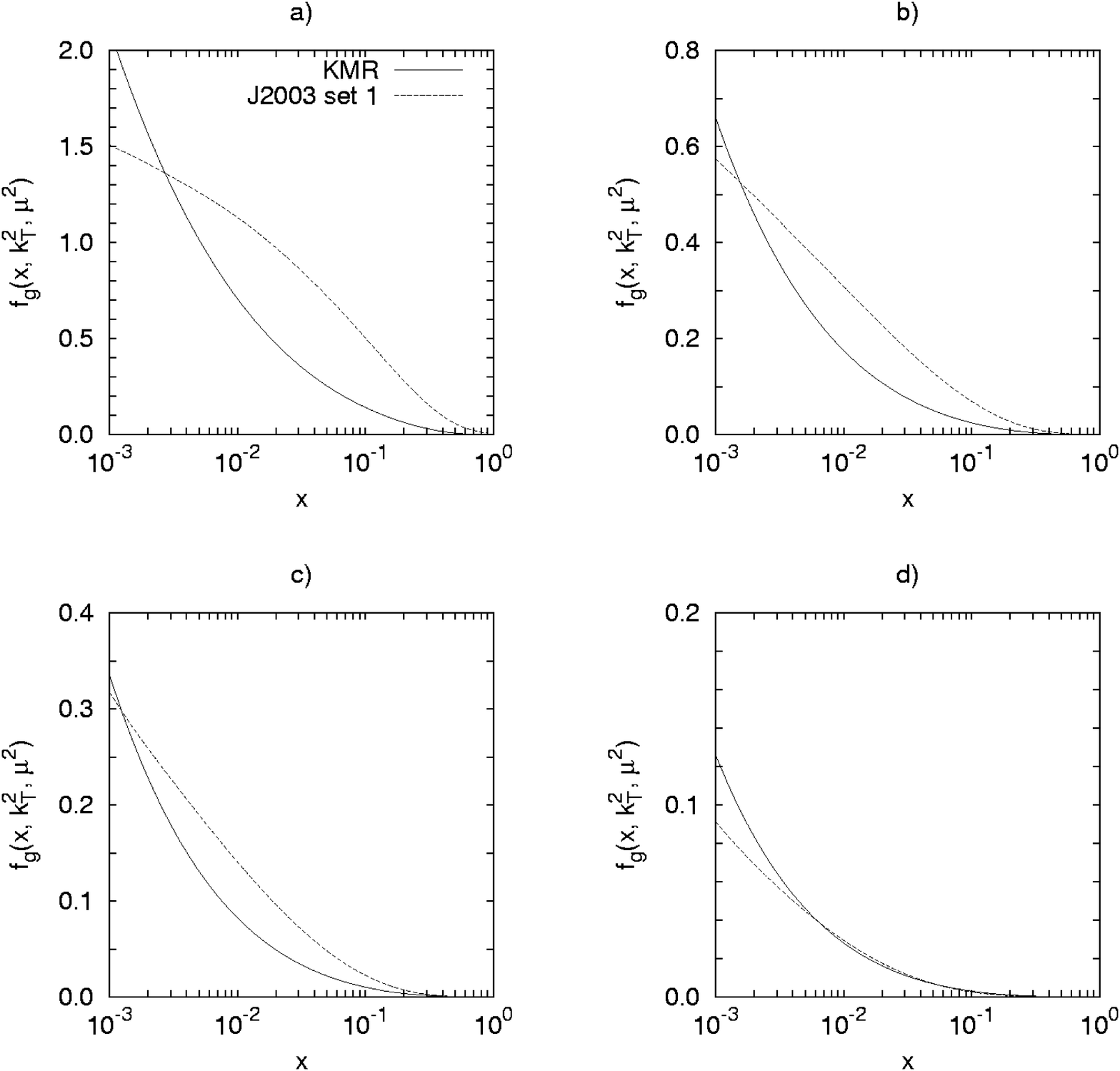, width = 17.5cm, height = 17.5cm}
\caption{The unitegrated gluon distributions $f_g(x,{\mathbf k}_T^2,\mu^2)$
at scale $\mu^2 = 100\,{\rm GeV}^2$ as a function 
of $x$ for different values of ${\mathbf k}_T^2$, namely 
${\mathbf k}_T^2 = 2\,{\rm GeV}^2$ (a),
${\mathbf k}_T^2 = 10\,{\rm GeV}^2$ (b),
${\mathbf k}_T^2 = 20\,{\rm GeV}^2$ (c) and 
${\mathbf k}_T^2 = 50\,{\rm GeV}^2$ (d).
The solid lines and dashed lines correspond to the KMR and J2003 set 1 gluon densities, 
respectively.}
\label{fig3}
\end{figure}

\newpage

\begin{figure}
\epsfig{figure=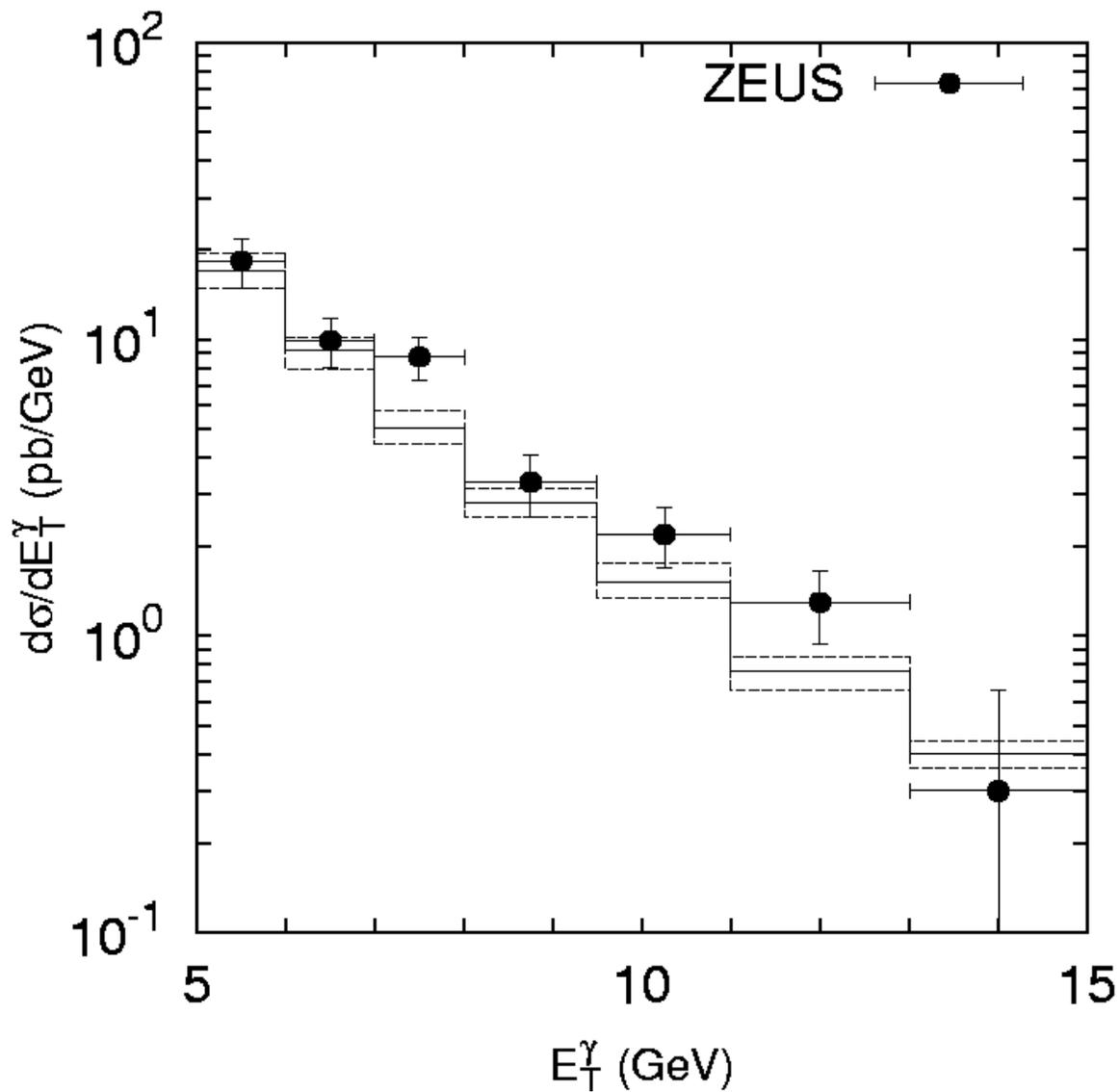, width = 22cm}
\caption{The differential cross section $d\sigma/d E_T^\gamma$ for the 
inclusive prompt photon production calculated at $ - 0.7 < \eta^\gamma < 0.9$ and $0.2 < y < 0.9$. 
The solid histogram corresponds to the default scale $\mu = E_T^\gamma$, whereas upper and 
lower dashed histograms correspond to the $\mu = E_T^\gamma/2$ and $\mu = 2 E_T^\gamma$ 
scales, respectively. The experimental data are from ZEUS~[3].}
\label{fig4}
\end{figure}

\newpage

\begin{figure}
\epsfig{figure=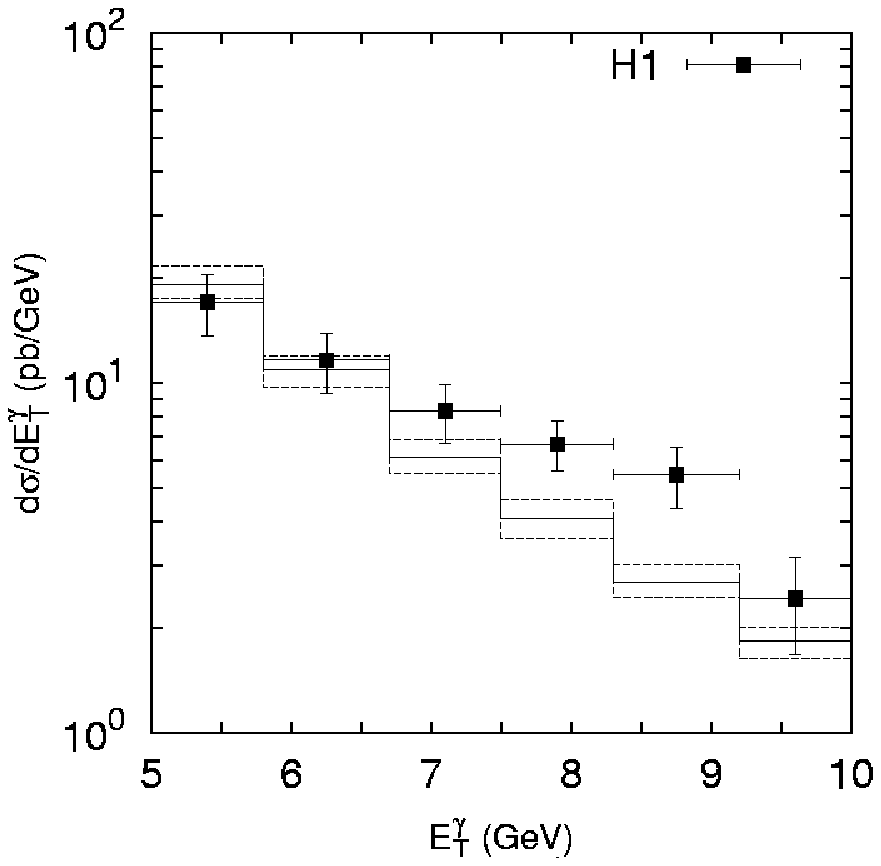, width = 22cm}
\caption{The differential cross section $d\sigma/d E_T^\gamma$ for the 
inclusive prompt photon production calculated at $ - 1 < \eta^\gamma < 0.9$ and $0.2 < y < 0.7$. 
All histograms are the same as in Figure~4. The experimental data are from H1~[4].}
\label{fig5}
\end{figure}

\newpage

\begin{figure}
\epsfig{figure=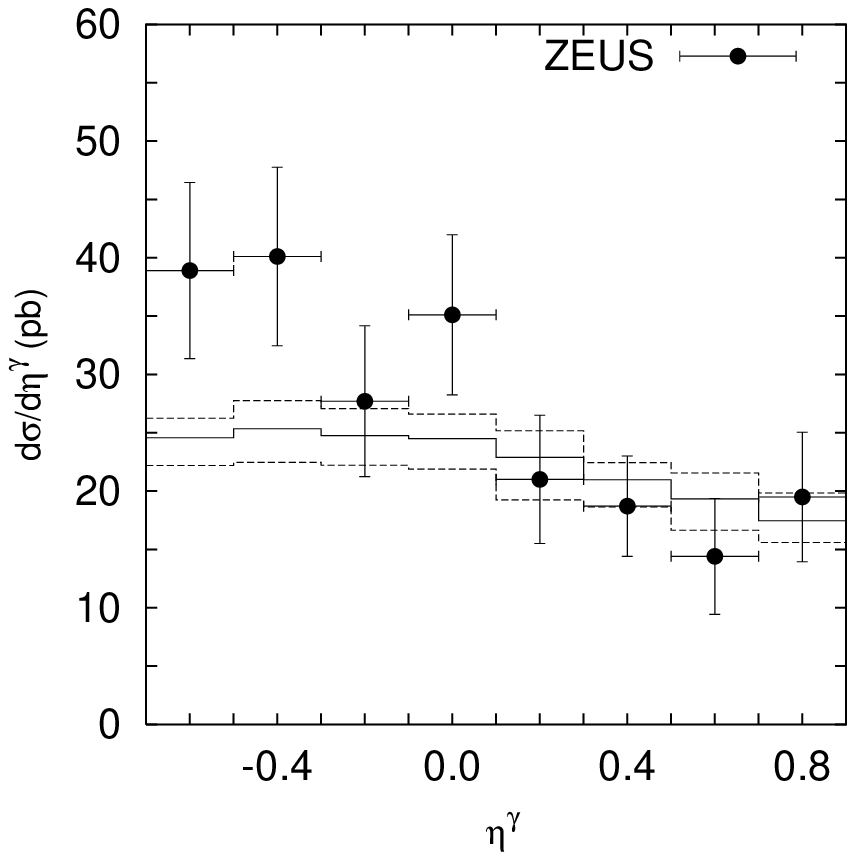, width = 22cm}
\caption{The differential cross section $d\sigma/d\eta^\gamma$ for the 
inclusive prompt photon production calculated at $5 < E_T^\gamma < 10$ GeV and $0.2 < y < 0.9$
($134 < W < 285$ GeV). All histograms are the same as in Figure~4. The experimental data 
are from ZEUS~[3].}
\label{fig6}
\end{figure}

\newpage

\begin{figure}
\epsfig{figure=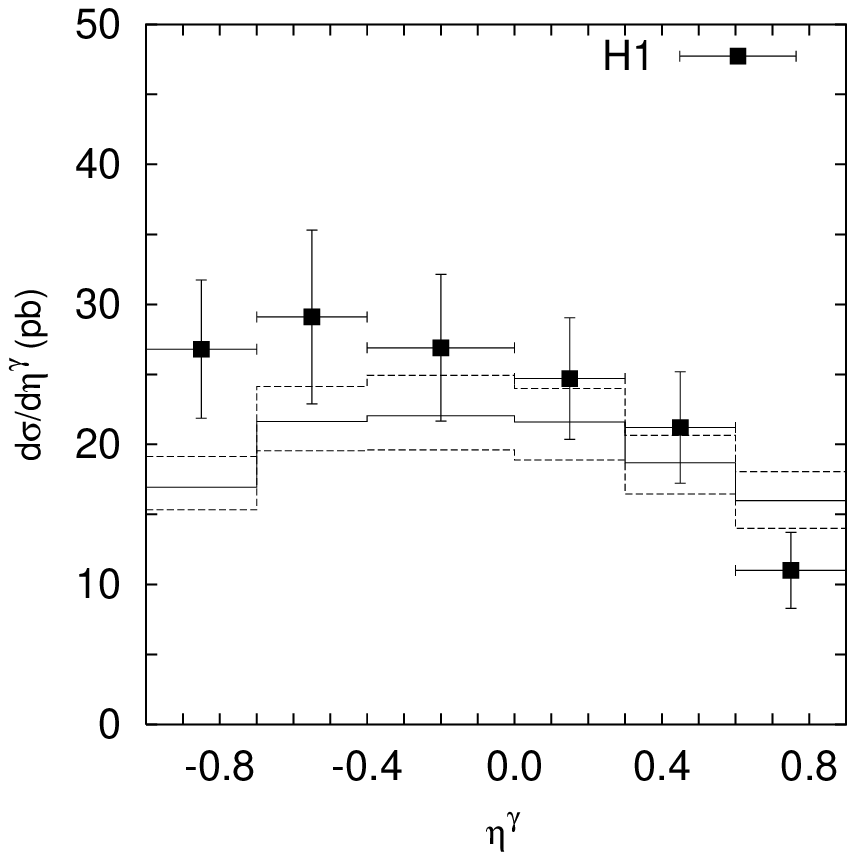, width = 22cm}
\caption{The differential cross section $d\sigma/d\eta^\gamma$ for the 
inclusive prompt photon production calculated at $5 < E_T^\gamma < 10$ GeV and $0.2 < y < 0.7$. 
All histograms are the same as in Figure~4. The experimental data are from H1~[4].}
\label{fig7}
\end{figure}

\newpage

\begin{figure}
\epsfig{figure=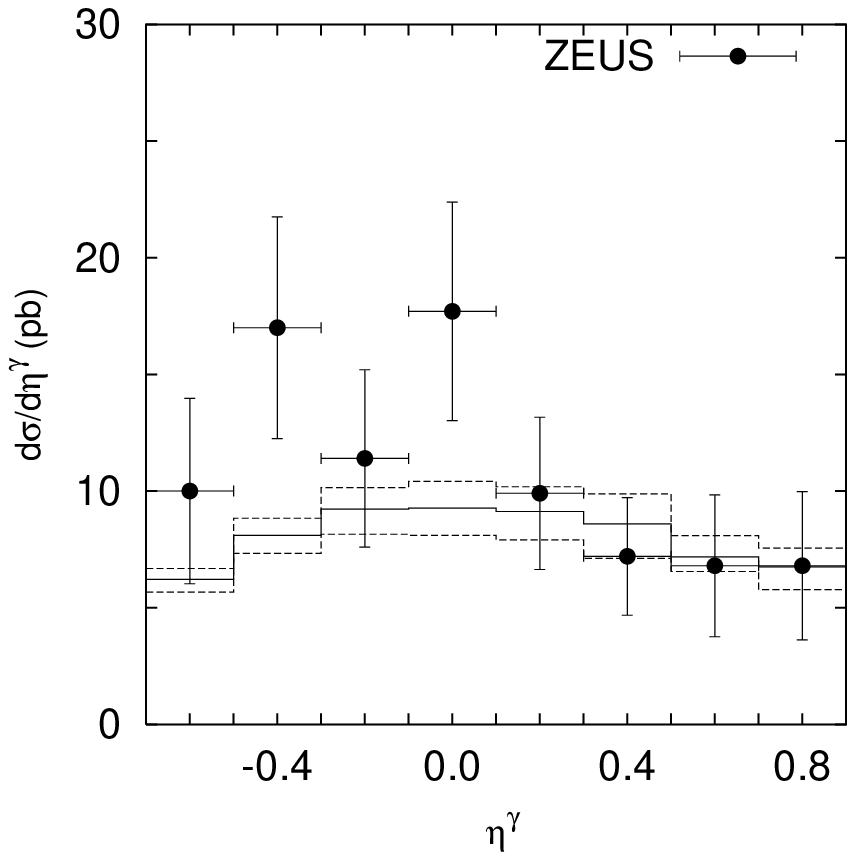, width = 22cm}
\caption{The differential cross section $d\sigma/d\eta^\gamma$ for the 
inclusive prompt photon production calculated at $5 < E_T^\gamma < 10$ GeV and $0.2 < y < 0.32$
($134 < W < 170$ GeV). All histograms are the same as in Figure~4. The experimental data 
are from ZEUS~[3].}
\label{fig8}
\end{figure}

\newpage

\begin{figure}
\epsfig{figure=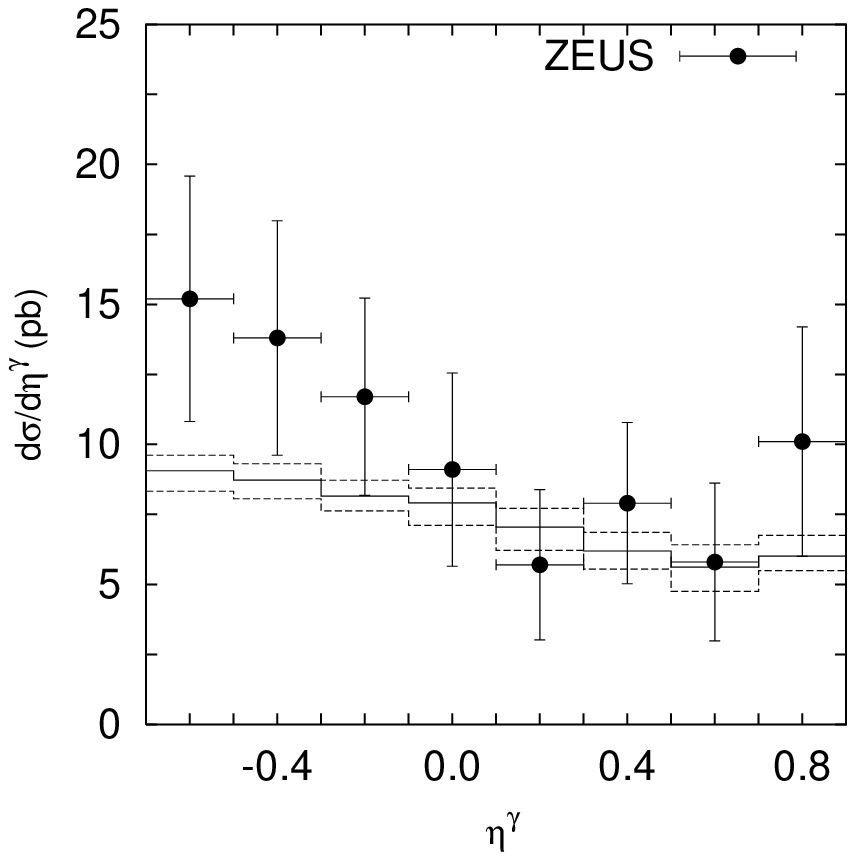, width = 22cm}
\caption{The differential cross section $d\sigma/d\eta^\gamma$ for the 
inclusive prompt photon production calculated at $5 < E_T^\gamma < 10$ GeV and $0.32 < y < 0.5$
($170 < W < 212$ GeV). All histograms are the same as in Figure~4. The experimental data 
are from ZEUS~[3].}
\label{fig9}
\end{figure}

\newpage

\begin{figure}
\epsfig{figure=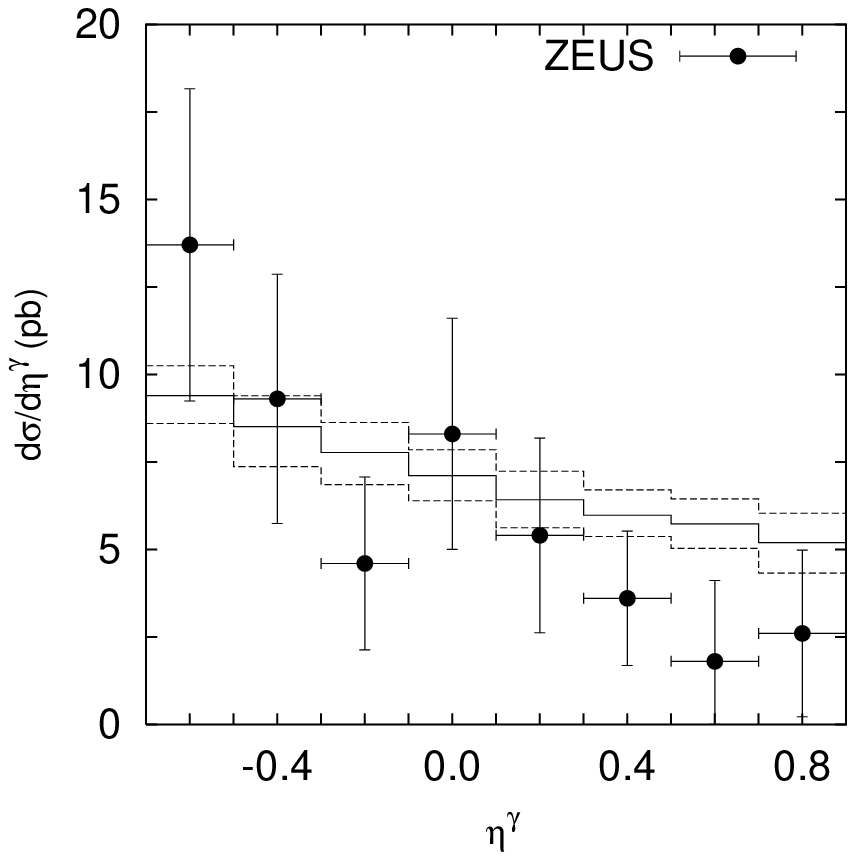, width = 22cm}
\caption{The differential cross section $d\sigma/d\eta^\gamma$ for the 
inclusive prompt photon production calculated at $5 < E_T^\gamma < 10$ GeV and $0.5 < y < 0.9$
($212 < W < 285$ GeV). All histograms are the same as in Figure~4. The experimental data 
are from ZEUS~[3].}
\label{fig10}
\end{figure}

\newpage

\begin{figure}
\epsfig{figure=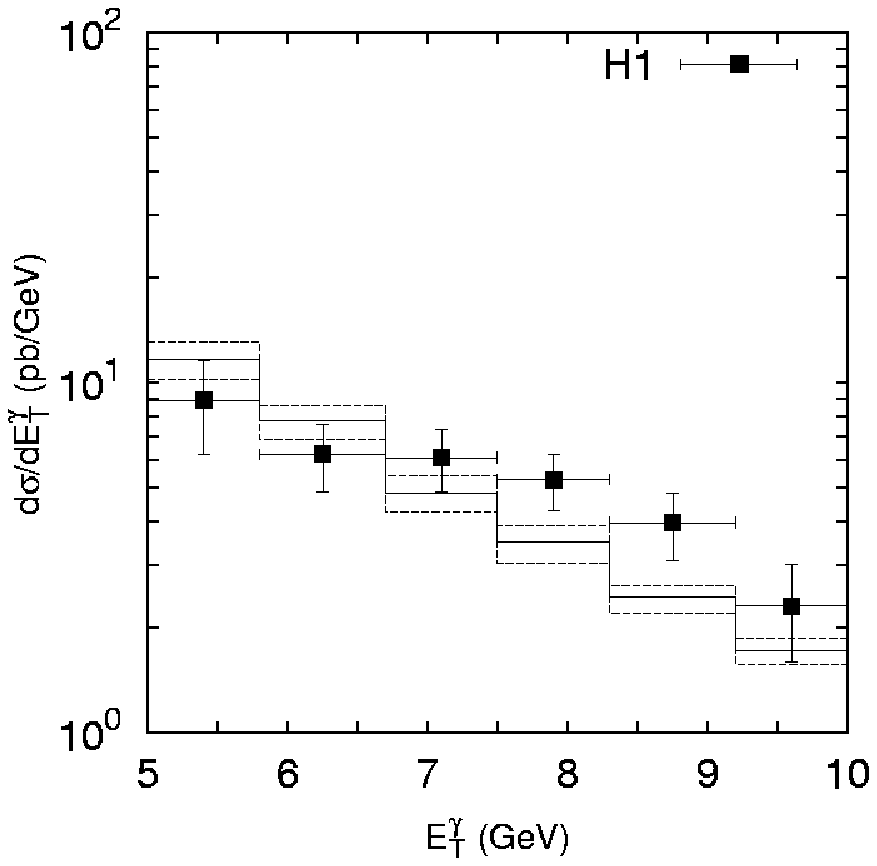, width = 22cm}
\caption{The differential cross section $d\sigma/d E_T^\gamma$ for the 
prompt photon + jet production calculated at $ - 1 < \eta^\gamma < 0.9$ and $0.2 < y < 0.7$
with an additional jet requirement $ - 1 < \eta^{\rm jet} < 2.3$ and $E_T^{\rm jet} > 4.5$ GeV.
All histograms are the same as in Figure~4. The experimental data are from H1~[4].}
\label{fig11}
\end{figure}

\newpage

\begin{figure}
\epsfig{figure=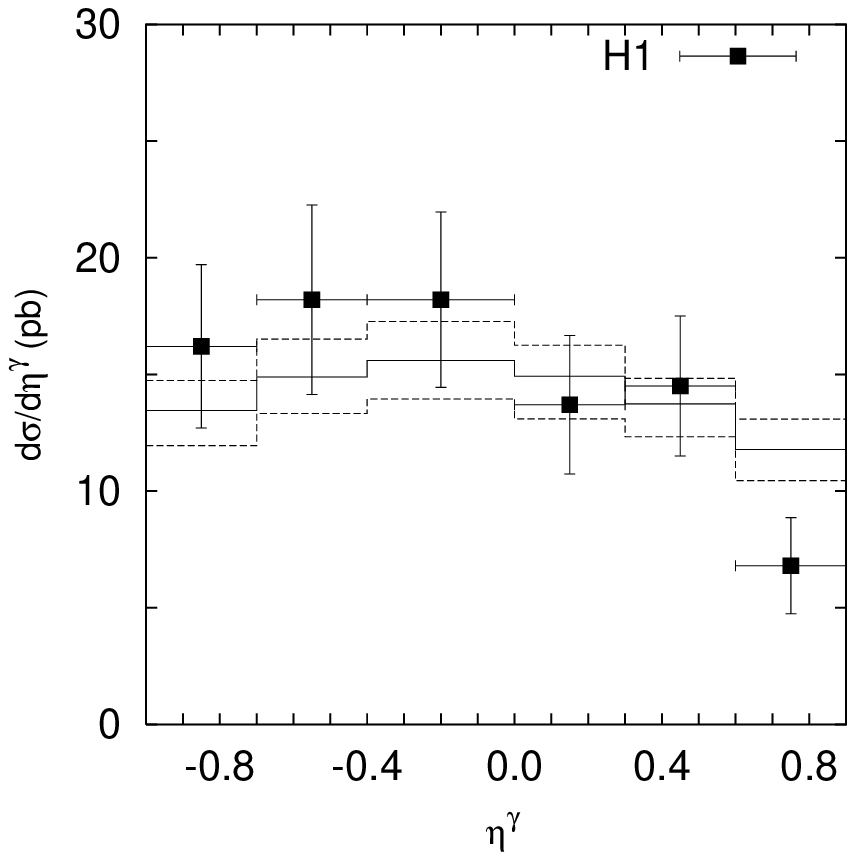, width = 22cm}
\caption{The differential cross section $d\sigma/d\eta^\gamma$ for the 
prompt photon + jet production calculated at $5 < E_T^\gamma < 10$ GeV and $0.2 < y < 0.7$
with an additional jet requirement $ - 1 < \eta^{\rm jet} < 2.3$ and $E_T^{\rm jet} > 4.5$ GeV.
All histograms are the same as in Figure~4. The experimental data are from H1~[4].}
\label{fig12}
\end{figure}

\newpage

\begin{figure}
\epsfig{figure=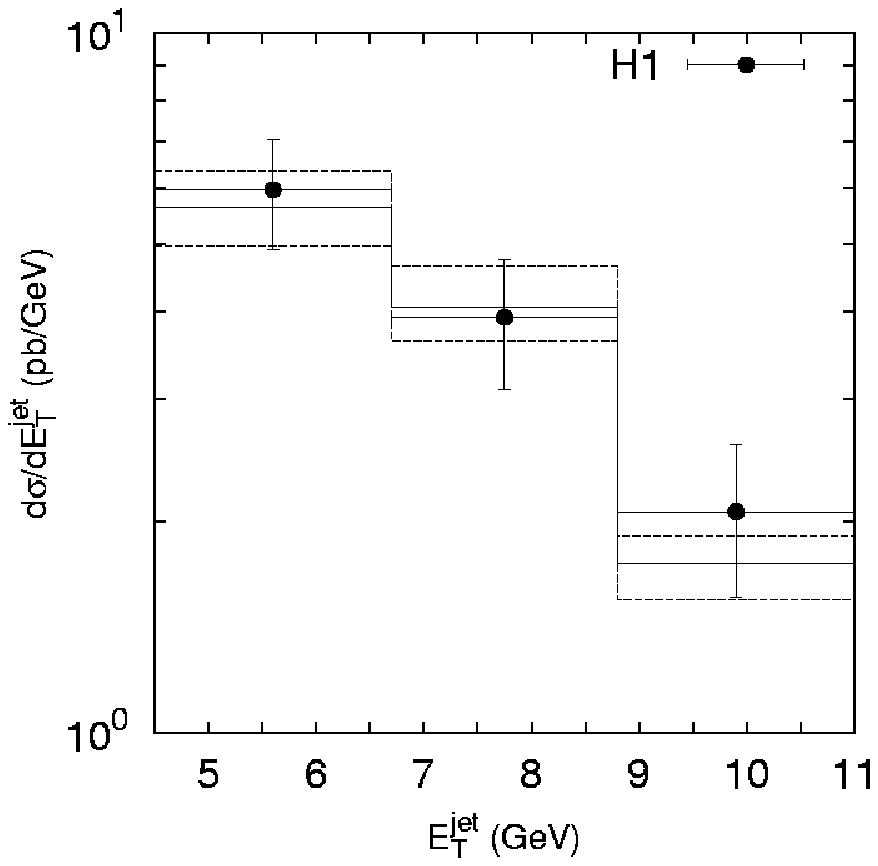, width = 22cm}
\caption{The differential cross section $d\sigma/dE_T^{\rm jet}$ for the 
prompt photon + jet production calculated at $5 < E_T^\gamma < 10$ GeV and $0.2 < y < 0.7$
with an additional jet requirement $ - 1 < \eta^{\rm jet} < 2.3$ and $E_T^{\rm jet} > 4.5$ GeV.
All histograms are the same as in Figure~4. The experimental data are from H1~[4].}
\label{fig13}
\end{figure}

\newpage

\begin{figure}
\epsfig{figure=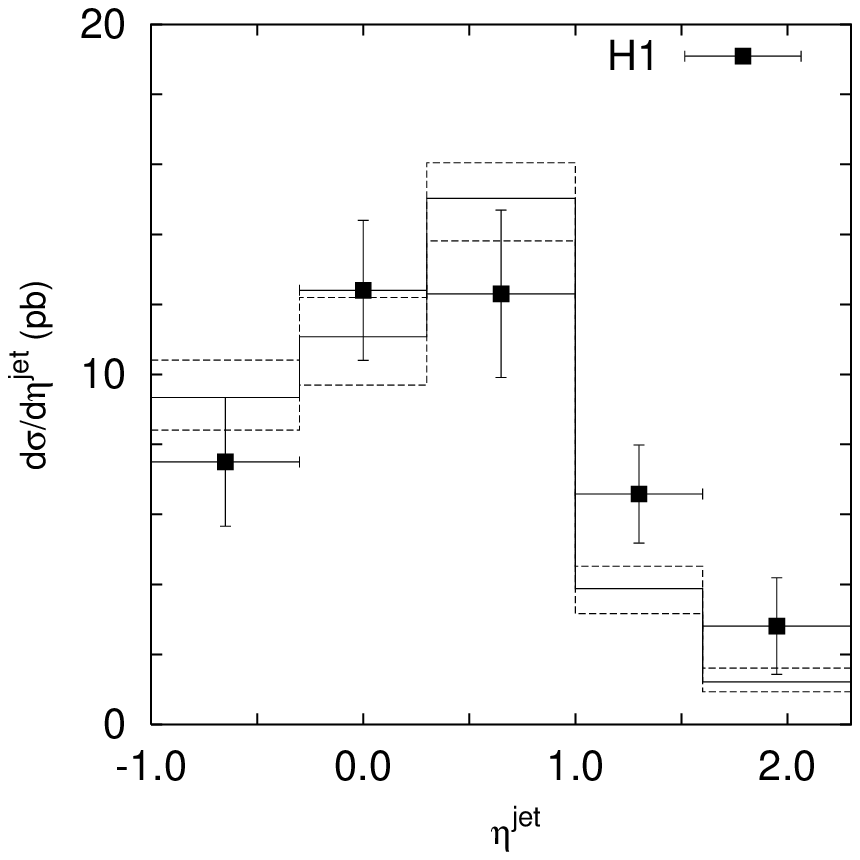, width = 22cm}
\caption{The differential cross section $d\sigma/d\eta^{\rm jet}$ for the 
prompt photon + jet production calculated at $5 < E_T^\gamma < 10$ GeV and $0.2 < y < 0.7$
with an additional jet requirement $ - 1 < \eta^{\rm jet} < 2.3$ and $E_T^{\rm jet} > 4.5$ GeV.
All histograms are the same as in Figure~4. The experimental data are from H1~[4].}
\label{fig14}
\end{figure}

\newpage

\begin{figure}
\epsfig{figure=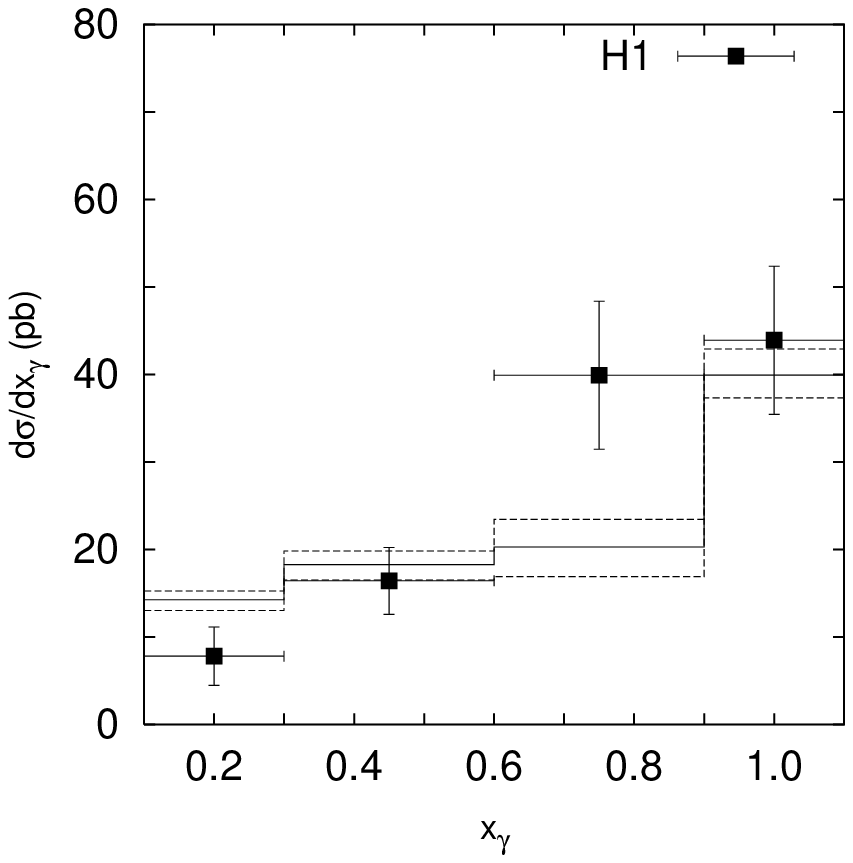, width = 22cm}
\caption{The differential cross section $d\sigma/dx_\gamma$ for the 
prompt photon + jet production calculated at $5 < E_T^\gamma < 10$ GeV and $0.2 < y < 0.7$
with an additional jet requirement $ - 1 < \eta^{\rm jet} < 2.3$ and $E_T^{\rm jet} > 4.5$ GeV.
All histograms are the same as in Figure~4. The experimental data are from H1~[4].}
\label{fig15}
\end{figure}

\newpage

\begin{figure}
\epsfig{figure=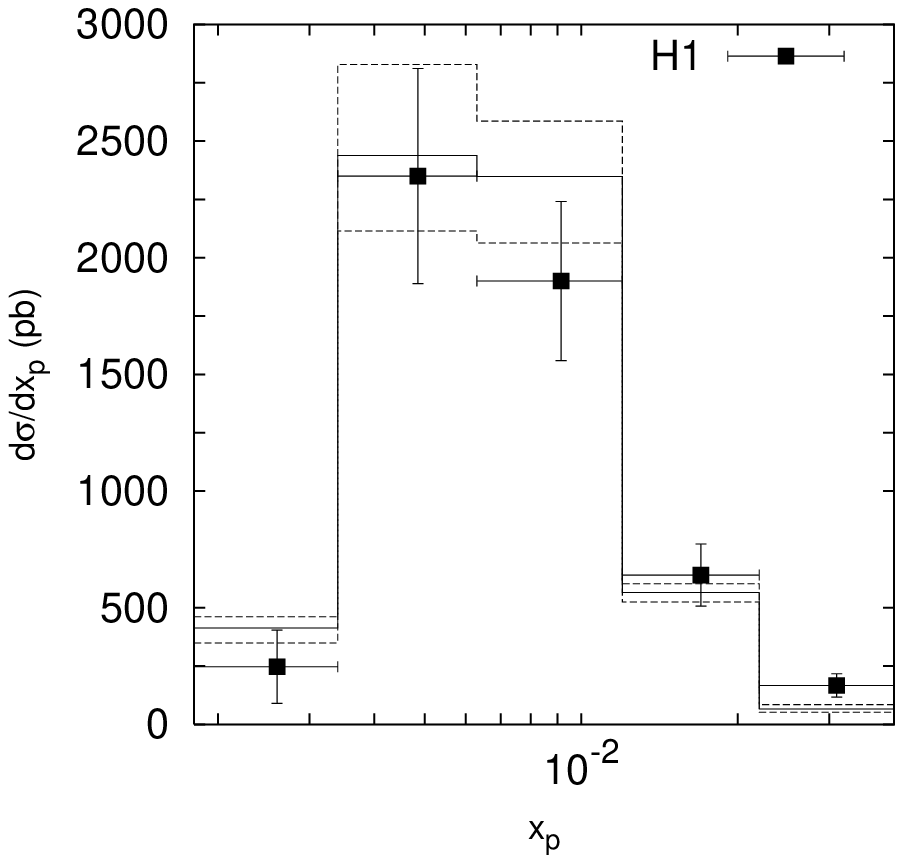, width = 21cm}
\caption{The differential cross section $d\sigma/dx_p$ for the 
prompt photon + jet production calculated at $5 < E_T^\gamma < 10$ GeV and $0.2 < y < 0.7$
with an additional jet requirement $ - 1 < \eta^{\rm jet} < 2.3$ and $E_T^{\rm jet} > 4.5$ GeV.
All histograms are the same as in Figure~4. The experimental data are from H1~[4].}
\label{fig16}
\end{figure}

\newpage

\begin{figure}
\epsfig{figure=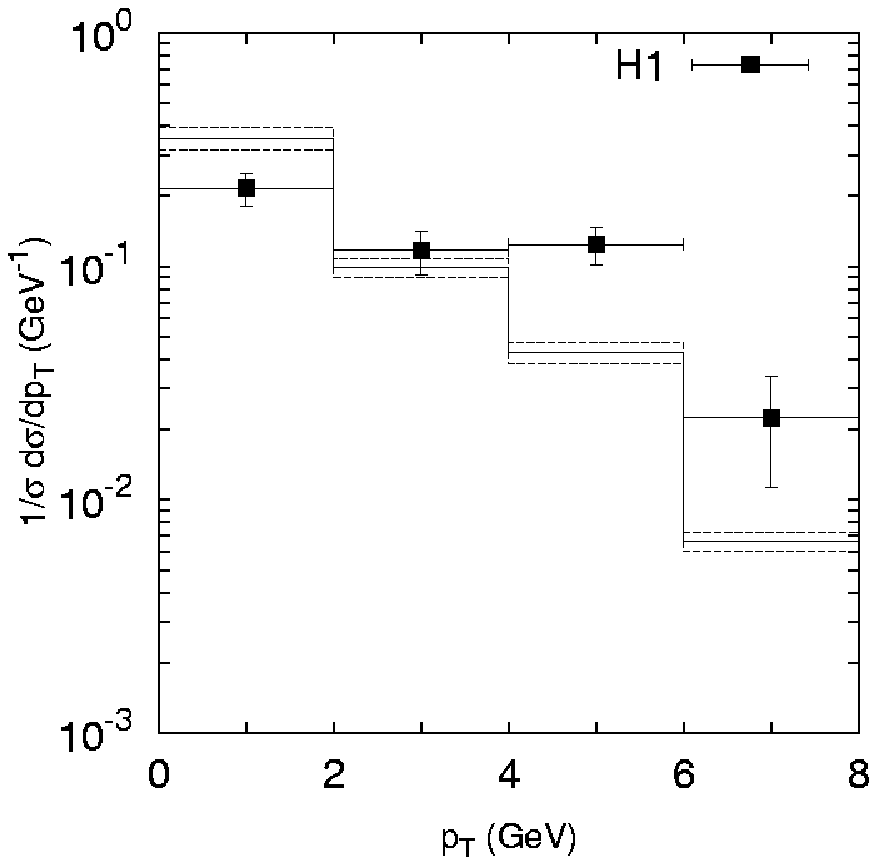, width = 22cm}
\caption{Distribution of the prompt photon momentum component, perpendicular
to the jet direction in the transverse plane at $5 < E_T^\gamma < 10$ GeV, $0.2 < y < 0.7$,
$ - 1 < \eta^{\rm jet} < 2.3$, $E_T^{\rm jet} > 4.5$ GeV and $x_\gamma < 0.85$.
All histograms are the same as in Figure~4. The experimental data are from H1~[4].}
\label{fig17}
\end{figure}

\begin{figure}
\epsfig{figure=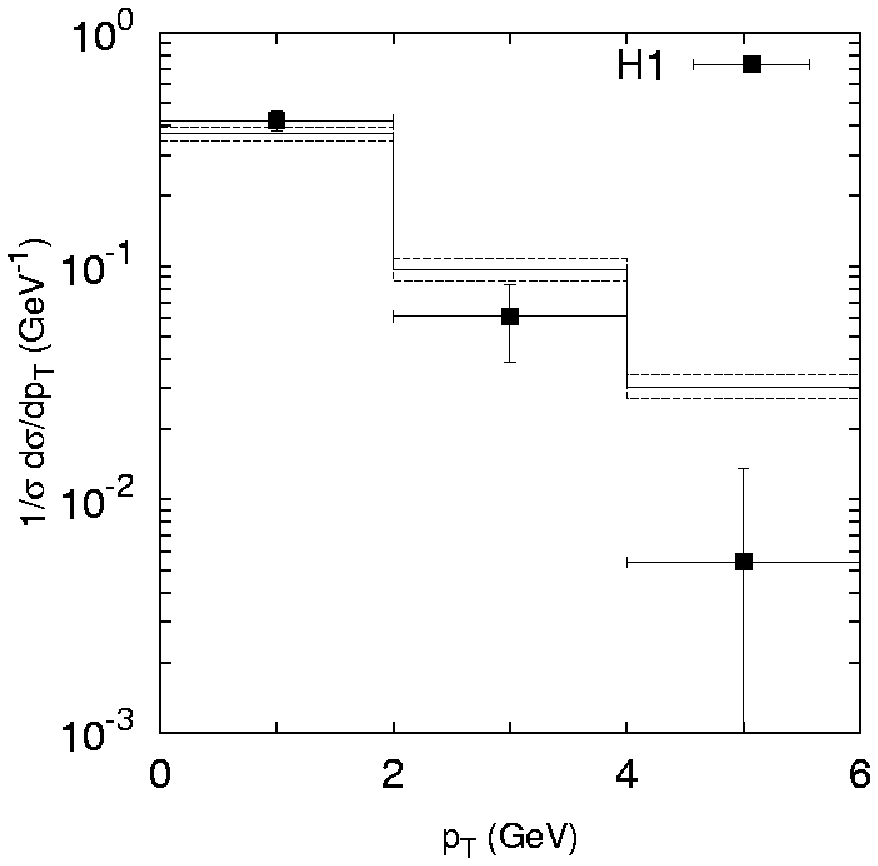, width = 22cm}
\caption{Distribution of the prompt photon momentum component, perpendicular
to the jet direction in the transverse plane at $5 < E_T^\gamma < 10$ GeV, $0.2 < y < 0.7$,
$ - 1 < \eta^{\rm jet} < 2.3$, $E_T^{\rm jet} > 4.5$ GeV and $x_\gamma > 0.85$.
All histograms are the same as in Figure~4. The experimental data are from H1~[4].}
\label{fig18}
\end{figure}

\end{document}